\documentclass[aps,prb,twocolumn,superscriptaddress,floatfix]{revtex4-2}
\usepackage{graphicx}
\usepackage{multirow}
\usepackage{color}
\usepackage{amsmath}
\usepackage{amsfonts}
\usepackage{amssymb}
\usepackage{graphicx}
\graphicspath{{Figures/}}
\usepackage{epstopdf}
\usepackage{empheq}
\usepackage{float}
\usepackage{cases} 
\usepackage{mathtools}
\usepackage{dcolumn}
\usepackage{bbold}
\usepackage{xfrac}
\usepackage{bm}
\usepackage{siunitx}

\usepackage[dvipsnames]{xcolor}
\usepackage[colorlinks]{hyperref}
\hypersetup{
colorlinks=True,
linkbordercolor=White,
linkcolor=BrickRed,
citecolor=Blue,	 
}

\DeclarePairedDelimiter\ket{\lvert}{\rangle}

\renewcommand{\vec}[1]{\mathbf{#1}}
\newcommand{\specialcell}[2][c]{%
\begin{tabular}[#1]{@{}c@{}}#2\end{tabular}}

\newcommand\YMN[1]{#1}
\newcommand\YMNN[1]{#1}

\begin{document}

\title{Geometry of the dephasing sweet spots of spin-orbit qubits}

\author{Lorenzo Mauro}
\thanks{These authors equally contributed to the work.}
\author{Esteban A. Rodr\'iguez-Mena}
\thanks{These authors equally contributed to the work.}
\affiliation{Univ. Grenoble Alpes, CEA, IRIG-MEM-L\_Sim, Grenoble, France.}
\author{Marion Bassi}
\author{Vivien Schmitt}
\affiliation{Univ. Grenoble Alpes, CEA, Grenoble INP, IRIG-Pheliqs, Grenoble, France.}%
\author{Yann-Michel Niquet}
\email{yniquet@cea.fr}
\affiliation{Univ. Grenoble Alpes, CEA, IRIG-MEM-L\_Sim, Grenoble, France.}%

\date{\today}

\begin{abstract}
The dephasing time of spin-orbit qubits is limited by the coupling with electrical and charge noise. However, there may exist ``dephasing sweet spots'' where the qubit decouples (to first order) from the noise so that the dephasing time reaches a maximum. Here we discuss the nature of the dephasing sweet spots of a spin-orbit qubit electrically coupled to some fluctuator. We characterize the Zeeman energy $E_\mathrm{Z}$ of this qubit by the tensor $G$ such that $E_\mathrm{Z}=\mu_B\sqrt{\vec{B}^\mathrm{T}G\vec{B}}$ (with $\mu_B$ the Bohr magneton and $\vec{B}$ the magnetic field), and its response to the fluctuator by the derivative $G^\prime$ of $G$ with respect to the fluctuating field. The geometrical nature of the sweet spots on the unit sphere describing the magnetic field orientation depends on the sign of the eigenvalues of $G^\prime$. We show that sweet spots usually draw lines on this sphere. We then discuss how to characterize the electrical susceptibility of a spin-orbit qubit with test modulations on the gates. We apply these considerations to a Ge/GeSi spin qubit heterostructure, and discuss the prospects for the engineering of sweet spots.
\end{abstract}

\maketitle

Hole spin qubits in semiconductor quantum dots \cite{Loss98,Burkard2023Review,Fang2023Review} have attracted much attention as possible building blocks for quantum computers and simulators \cite{Ares2013,Maurand16,Crippa18,Watzinger18,Hendrickx20,Hendrickx20b,Hendrickx21,Camenzind22,Froning21,wang2022ultrafast,Hendrickx2023,Wang2023}. They can indeed be manipulated electrically without extrinsic elements such as micro-magnets owing to the strong spin-orbit coupling (SOC) in the valence band of semiconductor materials \cite{Winkler03}. Moreover, the group IV elements such as silicon and germanium can be isotopically purified in order to get rid of the nuclear spins that may interact with the hole spins and spoil the coherence \cite{fischer2008spin,testelin2009hole,Tyryshkin12,Veldhorst14,Yoneda18,Mazzocchi19,moutanabbir23}. Nonetheless, the electrical addressability of such ``spin-orbit'' qubits comes along with a stronger sensitivity to electrical and charge noise.

It has however been demonstrated theoretically and experimentally that hole spin qubits can feature ``dephasing sweet spots'' as a function of the bias voltages and/or magnetic field orientation, where the Larmor frequency becomes insensitive (to first-order) to one or more source(s) of electrical noise \cite{Wang21,Bosco21,Malkoc2022,Piot22,michal2022tunable,Hendrickx2023}. The dephasing time $T_2^*$ then reaches a maximum near these sweet spots. For example, the echo time $T_2^{\mathrm E}$ of a hole spin qubit in a rectangular silicon channel could be extended up to $88$\,$\mu$s by selecting the magnetic field orientation where the Larmor frequency is least sensitive to the control gate voltages, hence to in-plane electric field fluctuations \cite{Piot22}. Interestingly, the optimal Rabi frequency (Rabi ```hot spot'') may also lie near one of the sweet spots thanks to reciprocal sweetness relations between dephasing (longitudinal spin susceptibility) and driving (transverse spin susceptibility) \cite{michal2022tunable}. This allows in principle a joint optimization of both the dephasing and manipulation times.

Charge traps in the amorphous gate oxide materials are believed to be the main source of electrical noise in semiconductor spin qubits \cite{Martinez2022,Kuppuswamy2022,Varley2023,massai2023,Shalak23}. This strengthened the interest in Ge/GeSi heterostructures, where the quantum dots are shaped by field effect in a thin Ge well embedded in thick expitaxial GeSi barriers \cite{Scappucci20,Hendrickx20,Hendrickx20b,Hendrickx21,Hendrickx2023}. The amorphous gate oxides are thus shifted at the surface of the heterostructure tens of nanometers away from the dots, which reduces the level of charge noise and disorder. Germanium hole spin qubits have actually seen outstanding progress in the past few years with the demonstration of a four qubit processor \cite{Hendrickx21} and of charge control in an array of sixteen dots \cite{Borsoi22}.

Whatever the host materials, it has become extremely important to understand the properties of the sweet spots of hole spin qubits in order to engineer dephasing times and achieve more resilient designs. In this work, we address the nature of these sweet spots from a general perspective. We demonstrate that the sweet spots of a given electrical fluctuator are most often ``sweet lines'' on the unit sphere describing the magnetic field orientation, as evidenced in Refs.~\cite{Piot22,michal2022tunable,Shalak23}. We also argue why the measurement of the derivative of the Larmor frequency with respect to the gate voltages provides a valuable assessment of the robustness of the qubit to electrical noise and helps locate operational sweet spots. Finally, we model a germanium hole spin qubit in a Ge/GeSi heterostructure \cite{martinez2022hole,Abadillo2023,Rodriguez2023} similar to a recent experiment \cite{Hendrickx2023} as an illustration. We discuss the nature and location of the sweet spots as a function of the symmetry (circular, squeezed \cite{Bosco21b}...) of the dot. We also analyse the impact of the inhomogeneous strains resulting from the contraction of the metal gates upon cooldown \cite{Abadillo2023,Corley2023,Liles20}.

The theory of sweet spots is discussed in section \ref{sec:sweetspots}, and the application to a Ge/GeSi spin qubit in section \ref{sec:applications}.

\section{Nature of the dephasing sweet spots}
\label{sec:sweetspots}

In this section, we discuss the dephasing sweet spots of a spin-orbit qubit electrically coupled to one or more fluctuators. We first briefly review the $g$-matrix formalism used for that purpose, \YMN{and introduce the longitudinal electric spin susceptibility (LSES) that characterizes the response of the Larmor frequency to electrical perturbations. We next analyze the geometrical nature of the dephasing sweet spots of a single fluctuator, then discuss the generalization to multiple fluctuators}. We finally address the experimental characterization of the LSES.

\subsection{Review of the $g$-matrix formalism}

In general, the effective Hamiltonian of a spin qubit in a homogeneous magnetic field $\vec{B}$ can be written
\begin{equation}
H=\frac{1}{2}\mu_B\boldsymbol{\sigma}\cdot g\vec{B}\,,
\label{eq:hamilt}
\end{equation}
where $\boldsymbol{\sigma}=(\sigma_1,\sigma_2,\sigma_3)$ are the Pauli matrices in a given $\{\ket{\Uparrow},\ket{\Downarrow}\}$ basis set for the two-level subspace, and $g$ is the gyromagnetic matrix \cite{Abragam1970,Crippa18,Venitucci18}. We emphasize that $g$ depends on the choice of the $\{\ket{\Uparrow},\ket{\Downarrow}\}$ basis set and of the $x$, $y$, $z$ axes for the magnetic field. It is not, in general, symmetric, but can always be factorized as
\begin{equation}
g=U{\tilde g}V^\mathrm{T}\,,
\end{equation}
where ${\tilde g}=\mathrm{diag}(g_1,g_2,g_3)$ is a diagonal matrix whose elements are the ``principal'' $g$-factors $g_1$, $g_2$, and $g_3$, and $U$, $V$ are real unitary matrices with determinants $\det U=\det V=+1$ \footnote{This factorization can be achieved, e.g., with a singular value decomposition. The constraint $\det U=\det V=+1$ is applied to get right-handed axes $\{\vec{v}_1,\,\vec{v}_2,\,\vec{v}_3\}$ and to associate $U$ with a rotation $R(U)$ in the two-level subspace along note 38 of Ref.~\cite{Venitucci18}). The sign of either $\det U$ or $\det V$ can be changed by reversing the sign of a principal $g$-factor $g_i$ and of the corresponding column of either $U$ or $V$. The signs of both $\det U$ and $\det V$ can be changed by reversing the sign of the same column of $U$ and $V$, or by swapping two principal $g$-factors $g_i$ as well as the corresponding columns of $U$ and $V$.}. The columns of $V$ define the principal magnetic axes $\vec{v}_1$, $\vec{v}_2$ and $\vec{v}_3$, while $U$ can be associated with a rotation $R(U)$ in the two-level subspace, thus with two basis states $\ket{\tilde\Uparrow}=R(U)\ket{\Uparrow}$ and $\ket{\tilde\Downarrow}=R(U)\ket{\Downarrow}$ \cite{Venitucci18}. With $(B_1,\,B_2,\,B_3)$ the coordinates of the magnetic field in the $\{\vec{v}_1,\,\vec{v}_2,\,\vec{v}_3\}$ axes, the Hamiltonian simply reads in the $\{\ket{\tilde\Uparrow},\ket{\tilde\Downarrow}\}$ basis set:
\begin{equation}
H=\frac{1}{2}\mu_B(g_1B_1\sigma_1+g_2B_2\sigma_2+g_3B_3\sigma_3)\,.
\end{equation}
More generally, a change of axes for the magnetic field results in a transformation $g\to gP^\mathrm{T}$, and a change of two-level basis set in a transformation $g\to Qg$, with $P$ and $Q$ real unitary matrices \cite{Venitucci18}.

The Hamiltonian~\eqref{eq:hamilt} can also be written
\begin{equation}
H=\frac{1}{2}\mu_Bg^*B\sigma^*\,,
\end{equation}
where $B=|\vec{B}|$, $g^*=|g\vec{b}|$ is the effective gyromagnetic factor for a magnetic field oriented along the unit vector $\vec{b}=\vec{B}/B$, and $\sigma^*=\boldsymbol{\sigma}\cdot\vec{u}$ is the matrix of an effective spin along the unit vector $\vec{u}=g\vec{b}/g^*$. The eigenenergies are therefore
\begin{equation}
E_\pm=\pm\frac{1}{2}g^*\mu_BB
\end{equation}
and the Zeeman splitting is $E_\mathrm{Z}=g^*\mu_BB$. We can thus introduce the Zeeman tensor
\begin{equation}
G=g^\mathrm{T}g   
\end{equation}
such that $g^*=\sqrt{\vec{b}^\mathrm{T}G\vec{b}}$ \cite{Crippa18,Venitucci18}. Contrary to $g$, $G$ is always symmetric, and depends only on the choice of axes for the magnetic field (as a transformation $g\to Qg$ leaves $G$ invariant). Its eigenvalues are the squares $g_i^2$ of the principal $g$-factors, and its eigenvectors are the principal magnetic axes $\{\vec{v}_1,\,\vec{v}_2,\,\vec{v}_3\}$:
\begin{equation}
G=V({\tilde g}^2)V^\mathrm{T}\,.
\end{equation}
The effective gyromagnetic factor is hence simply $g^*=\sqrt{g_1^2b_1^2+g_2^2b_2^2+g_3^2b_3^2}$ in the principal magnetic axes. The Zeeman tensor is an observable that can be constructed from a measurement of the Zeeman splitting in at least six orientations of the magnetic field (since there are six independent matrix elements in $G$). The sign of the principal $g$-factors may, nonetheless, remain ambiguous as the diagonalization of $G$ only provides the $g_i^2$'s. As discussed below, the dephasing rate of a spin qubit electrically coupled to some noise can be expressed as a function of $G$ and its derivative with respect to the noisy parameter. The $g$ matrix itself (and more so its derivatives) are much more difficult to reconstruct from experimental data \cite{Crippa18}, as the working two-level basis set $\{\ket{\Uparrow},\ket{\Downarrow}\}$ can hardly be made explicit. The $g$ matrix is, however, a very useful asset for modeling, as it can be easily computed from first-order perturbation theory once states $\ket{\Uparrow}$ and $\ket{\Downarrow}$ have been computed \cite{Venitucci18}.  

\subsection{The longitudinal spin susceptibilities}

Let us now introduce a fluctuator characterized by some stationary random signal $A(t)$ (a fluctuating gate voltage, charge or dipole for example). \YMN{This fluctuator can induce decoherence \cite{Paladino14} through relaxation and dephasing (usually dominant in spin qubits and the focus of this work \footnote{\YMN{The random signal $A(t)$ may indeed induce transitions between the spin states and thus contribute to the relaxation rate $\Gamma_1=1/T_1$ \cite{Paladino14}. This contribution to $\Gamma_1$ is, to lowest order, proportional to the power spectrum $S_A(f_\mathrm{L})$ of $A(t)$ at the Larmor frequency $f_\mathrm{L}$ [whereas the dephasing rate is most sensitive to the low-frequency tail $S_A(f\ll f_\mathrm{L})$]. We assume throughout this work that decoherence is dominated by dephasing (which is usually the case in spin qubits coupled to $1/f$ noise).}}). Dephasing results from the modulations of the Larmor frequency $f_\mathrm{L}(A)=E_\mathrm{Z}(A)/h$ by the fluctuator (with $h$ the Planck constant). The phase accumulated over time $t$ is indeed:}
\begin{equation}
\Phi(t)=2\pi\int_0^t\,dt^\prime f_\mathrm{L}(A(t^\prime))=2\pi f_\mathrm{L}(0)t+\Delta\Phi(t),
\end{equation}
where the phase shift $\Delta\Phi(t)$ with respect to free precession reads, to first order in $A$:
\begin{equation}
\Delta\Phi(t)=2\pi f_\mathrm{L}^\prime(0)\times\int_0^t\,dt^\prime A(t^\prime)\,.
\end{equation}
This phase shift is therefore proportional to $f_\mathrm{L}^\prime=df_\mathrm{L}/dA$, the longitudinal spin susceptibility (LSS) with respect to the perturbation $A$. As an illustration, we consider random signals $A(t)$ with a power spectrum $S_A(f)\propto 1/f$ over a frequency bandwidth $f\in [f_\mathrm{min},f_\mathrm{max}]$ (but the theory can easily be extended to other classes of noises). The coherence then decays as $\exp[-(t/T_2^*)^2]$ when $t\ll 1/f_\mathrm{max}$ \cite{Ithier2005}, with the pure dephasing time $T_2^*$ given by 
\begin{equation}
\Gamma_2^*=\frac{1}{T_2^*}=\sqrt{2}\pi A^\mathrm{rms}|f_\mathrm{L}^\prime(0)|
\label{eq:Gamma2}
\end{equation}
and $A^\mathrm{rms}$ the rms fluctuations of $A(t)$ \YMN{\footnote{\YMN{This expression provides a fair account of the contribution of $1/f$ noise to decoherence when $\Gamma_2^*\gg f_\mathrm{max}$ (since it only holds for ``short'' times $t\ll 1/f_\mathrm{max}$). There are, in particular, logarithmic corrections beyond this regime. For a complete discussion, see for example Ref.~\cite{Ithier2005}.}}}. The LSS $f_\mathrm{L}^\prime(0)$ and $A^\mathrm{rms}$ hence completely characterize the dephasing rate $\Gamma_2^*$ to first order in $A$. 

The LSS may result from modulations of the Zeeman tensor $G$ (for electrically coupled fluctuators) or from modulations of the magnetic field $\vec{B}$ (for magnetically coupled fluctuators). In hole spin qubits, most relevant fluctuators primarily couple electrically to the hole charge, then to the hole spin through SOC (one exception being nuclear spins, however absent in isotopically purified materials \cite{fischer2008spin,testelin2009hole,moutanabbir23}). The LSS is then a longitudinal spin electric susceptibility (LSES), which can be related to the Zeeman tensor $G$ and its derivative $G^\prime=dG/dA$:
\begin{equation}
f_\mathrm{L}^\prime=\frac{\mu_B B}{2h}\frac{\mathbf{b}^\mathrm{T}G^\prime(0)\mathbf{b}}{\sqrt{\mathbf{b}^{\mathrm{T}}G(0)\mathbf{b}}}=\frac{\mu_B B}{2hg^*}\mathbf{b}^\mathrm{T}G^\prime(0)\mathbf{b}\,.
\label{eq:lses}
\end{equation}
Note that $G^\prime$ is also symmetric by design. $G^\prime(0)$ can, therefore, be diagonalized and factorized as
\begin{equation}
G^\prime(0)=W\tilde{G}^\prime W^\mathrm{T}
\end{equation}
where the elements of $\tilde{G}^\prime=\mathrm{diag}(G_1^\prime,G_2^\prime,G_3^\prime)$ are the eigenvalues of $G^\prime$ and the columns $\{\vec{w}_1,\,\vec{w}_2,\,\vec{w}_3\}$ of $W$ are the corresponding eigenvectors (that may differ from $\vec{v}_1$, $\vec{v}_2$, and $\vec{v}_3$).

For the purpose of analysis, we can further split $G^\prime=\Gamma^\prime+\Xi^\prime$ in two matrices:
\begin{subequations}
\label{eq:gammaxi}
\begin{align}
\Gamma^\prime&=V\frac{\partial\tilde{g}^2}{\partial A}V^\mathrm{T} \\
\Xi^\prime&=\frac{\partial V}{\partial A}\tilde{g}^2V^\mathrm{T}+\mathrm{transpose}\,.\label{eq:xi}
\end{align}
\end{subequations}
$\Gamma^\prime$ accounts for the modulations of the principal $g$-factors $g_1$, $g_2$ and $g_3$, and $\Xi^\prime$ for the modulations of the principal axes $\vec{v}_1$, $\vec{v}_2$, and $\vec{v}_3$. Moreover, since $V$ is a unitary matrix, 
\begin{equation}
\frac{\partial}{\partial A}(V^\mathrm{T}V)=0=V^\mathrm{T}\left(\frac{\partial V}{\partial A}\right)+\mathrm{transpose}\,,
\end{equation}
so that the matrix
\begin{equation}
Z=V^\mathrm{T}\left(\frac{\partial V}{\partial A}\right)\,,
\end{equation}
\YMN{which is independent on the choice of magnetic axes}, is antisymmetric ($Z^\mathrm{T}=-Z$) and has thus diagonal elements $Z_{ii}=0$. It then follows from Eqs.~\eqref{eq:gammaxi} that:
\begin{equation}
\frac{\partial\tilde{g}^2}{\partial A}=\mathrm{diag}\left(G_V^\prime\right)\,,
\end{equation}
where $G_V^\prime=V^\mathrm{T}G^\prime V$ is the derivative of $G$ in the principal magnetic axes, and that:
\begin{equation}
Z_{ij}=\frac{(G_V^\prime)_{ij}}{g_j^2-g_i^2},\,i\ne j\,.
\end{equation}
This enables a full reconstruction of the derivatives of principal $g$-factors and principal magnetic axes from the measurement or calculation of $G^\prime$ (at least when the $g_i^2$'s are non degenerate, otherwise the principal magnetic axes of $G$ are not uniquely defined). As discussed in section \ref{sec:applications}, $G^\prime$ actually provides valuable information about the spin-orbit coupling mechanisms at work in the device.

\subsection{Dephasing sweet spots classification}
\label{sec:swclass}

There may exist particular bias points and/or orientations of the magnetic field where the LSS $f_\mathrm{L}'$ is zero and the qubit precession decouples (to first order) from the fluctuator. At these first-order ``sweet spots'', the pure dephasing time $T_2^*$ from Eq.~\eqref{eq:Gamma2} diverges; it gets practically limited by second or higher-order couplings to the noise and thus reaches a maximum in the close vicinity of the sweet spot if these residual couplings are small enough.

According to Eq.~\eqref{eq:lses}, the first-order sweet spots for electrically coupled fluctuators are the solutions of $\mathbf{b}^\mathrm{T}G^\prime(0)\mathbf{b}=0$. This defines a set of magnetic field orientations; the components $(b_1^\prime,\,b_2^\prime,\,b_3^\prime)$ of $\mathbf{b}$ in the principal axes $\{\vec{w}_1,\,\vec{w}_2,\,\vec{w}_3\}$ of $G^\prime(0)$ must actually satisfy:
\begin{subequations}
\begin{align}
&G_1^\prime b_1^{\prime 2}+G_2^\prime b_2^{\prime 2}+G_3^\prime b_3^{\prime 2}=0 \label{eq:sw1}\\
&b_1^{\prime 2}+b_2^{\prime 2}+b_3^{\prime 2}=1\,. \label{eq:sw2}
\end{align}
\end{subequations}
The sweet spot orientations are thus at the intersection between the quadric surface defined by Eq.~\eqref{eq:sw1} and the unit sphere. Depending on the signs of the eigenvalues $G_i^\prime$, this intersection can take different shapes, listed in Table~\ref{tab:sw}. The second column is the geometrical object defined by Eq.~\eqref{eq:sw1}, and the third one is the shape of its intersection with the unit sphere. There are no sweet spots if all $G_i^\prime$'s have the same sign; on the contrary there may be ``global'' sweet spots (independent on the magnetic field orientation) at the specific bias points where $G^\prime(0)$ is identically zero \cite{Bosco21,michal2022tunable}. For hole spin qubits, the $G_i^\prime$'s usually take different signs when the fluctuator reshapes the wave function, so that the sweet spots generally lie on one or a couple of lines, as evidenced in Refs. \cite{michal2022tunable,Piot22,Shalak23}. This leaves, in principle, more opportunities for optimal operation of the qubit. Whatever their geometry, the sweet spots must be invariant by the inversion $\vec{b}\to-\vec{b}$.

\begin{table}[t]
\noindent \begin{centering}
\begin{tabular}{|c|c|c|}
\hline 
Signs of the $G_i^\prime $ & \specialcell{Solutions of\\ $\mathbf{b}^\mathrm{T}G^\prime(0)\mathbf{b}=0$} & \specialcell{Sweet spot\\ orientations} \tabularnewline
\hline 
\hline 
$(0,\,0,\,0)$ & Whole space & Unit sphere\tabularnewline
\hline 
\specialcell{$(0,\,0,\,+)$\\ $(0,\,0,\,-)$} & 1 plane & 1 Sweet line\tabularnewline
\hline 
\specialcell{$(0,\,+,\,+)$\\ $(0,\,-,\,-)$} & 1 line & 2 Sweet points\tabularnewline
\hline 
$(0,\,+,\,-)$ & 2 planes & 2 Sweet lines\tabularnewline
\hline 
\specialcell{$(+,\,+,\,-)$\\ $(-,\,-,\,+)$} & 2 cones & 2 Sweet lines\tabularnewline
\hline 
\specialcell{$(+,\,+,\,+)$\\ $(-,\,-,\,-)$} & $\vec{b}=\vec{0}$ & None\tabularnewline
\hline 
\end{tabular}
\par\end{centering}
\label{tab:sw}
\caption{Geometrical nature of the sweet spot orientations as a function of the sign of the eigenvalues of $G^\prime(0)$ given in the first column. The second column is the geometrical nature of the solutions of Eq.~\eqref{eq:sw1}, and the third column the geometrical nature of their intersection with the unit sphere, which defines the sweet spot orientations.}
\end{table}

Moreover, we can express $G^\prime$ as a function of the matrices $g$ and $g^\prime$ computed in the same (but arbitrary) $\{\ket{\Uparrow},\ket{\Downarrow}\}$ basis set for the two-level subspace:
\begin{equation}
G^\prime=(g^\prime)^\mathrm{T}g+g^\mathrm{T}(g^\prime)\,.
\end{equation}
This provides a geometrical interpretation of the sweet spot condition $\mathbf{b}^\mathrm{T}G^\prime(0)\mathbf{b}=0$ \cite{michal2022tunable}. Indeed, a sweet spot  results from the achievement of either:
\begin{itemize}
\item $g^\prime(0)\vec{b}=\vec{0}$ (no modulations of $g$ for a particular orientation of the magnetic field).
\item $g(0)\vec{b}\perp g^\prime(0)\vec{b}$ (the modulations of the Larmor vector $\boldsymbol{\Omega}=g\vec{B}$ are orthogonal to the latter).
\end{itemize}
The second case is typical of sweet lines and the most usual in practice. We will further discuss its significance in section \ref{sec:recriprocity}. The case $g(0)\vec{b}=\vec{0}$ does not necessarily give rise to a sweet spot owing to the denominator of Eq.~\eqref{eq:lses} and is little relevant, since there is no Zeeman splitting.

Finally, we would like to emphasize that there is generally no exact sweet spot when there are two or more fluctuators in the vicinity of the qubit unless symmetries constrain the matrix $G^\prime$. Indeed, each fluctuator comes with its own sweet spots that may not coincide. The sweet lines of two fluctuators may still intersect at a few sweet points, but the intersection of $\gtrsim 3$ such lines is usually void. Nevertheless, the sweet spots of fluctuators located near the same symmetry elements remain usually close one to each other (e.g., the sweet spots of charge traps located under the same gate, as illustrated in Appendix \ref{app:traps}). The existence of  reliable ``quasi''-sweet spots with improved performance thus depends on material- and device-specific conditions such as the nature of the dominant noise and the broadness of the sweet features (with respect to the magnetic field orientation and bias voltages). This will be discussed in more detail for Ge/GeSi devices in section \ref{sec:applications}. \YMNN{In any case, the above considerations enable a meaningful characterization of each individual fluctuator, and an analysis of the prospects for design optimizations aiming to bring the sweet spots as close as possible to each other in order to maximize the coherence time.}

\subsection{The LSES of the gates as prototypical responses to the noise}

The electrostatics of a spin qubit is strongly constrained by the gate layout. The gates indeed shape the quantum dots, set the symmetries and pattern the electric field created by, e.g., charge defects. It is, therefore, useful to introduce the LSES of each gate
\begin{equation}
\mathrm{LSES}(\mathrm{G}_n)=\frac{\partial f_\mathrm{L}}{\partial V_{\mathrm{G}_n}}
\end{equation}
computed at the working bias point ($V_{\mathrm{G}_n}$ being the voltage on gate G$_n$). These LSES probe the response of the qubit to representative electrical perturbations with specific symmetries. They can be measured experimentally by monitoring the Larmor frequency as a function of gate voltages, and are thus very helpful in the exploration of the sweet spots and in the understanding of spin-orbit coupling in a device \cite{Piot22,Hendrickx2023}.

The dependence of the dephasing rates on the magnetic field orientation can even be possibly reconstructed from the LSES of the $N$ gates. Assuming again $1/f$-like noise with rms fluctuations $\delta V_{\mathrm{G}_n}^\mathrm{rms}$ on the different gates, the total dephasing rate reads:
\begin{equation}
\Gamma_2^*=\frac{1}{T_2^*}=\sqrt{2}\pi\sqrt{\sum_{n=1}^N\left(\delta V_{\mathrm{G}_n}^\mathrm{rms}\mathrm{LSES}(\mathrm{G}_n)\right)^2}\,.
\label{eq:gammatot}
\end{equation}
Each $\delta V_{\mathrm{G}_n}^\mathrm{rms}$ can be fitted to experimental data and must then be understood as an effective modulation that lumps the contributions from many fluctuators whose electric field shares close symmetry with that of the gate \footnote{Ensemble of fluctuators are actually a pre-requisite for $1/f$-like noise; the power spectrum of a single, telegraphic two-level fluctuator is Lorentzian \cite{Shalak23,Bergli09}.}. In Ref. \cite{Piot22}, the dephasing times of a silicon qubit were successfully analyzed along these lines. The long coherence achieved for a specific magnetic field orientation was, in particular, explained by the existence of nearby sweet spots for all gates. The relevance of Eq.~\eqref{eq:gammatot} depends on how far gate noise can mimic the main fluctuators. We further address this question on a Ge/GeSi spin qubit in section \ref{sec:applications} and in Appendix \ref{app:traps}. We emphasize that for $1/f$ noise $\Gamma_2^*$ is proportional, as are the LSES, to the magnetic field amplitude $B$ (the LSES must actually be zero when $B=0$ because an electric field can not couple to the spin if time-reversal symmetry is not broken).

\subsection{Quality factor}
\label{sec:recriprocity}
 
To go further, we can characterize the efficiency of single qubit gates by the quality factor:
\begin{equation}
Q_2^*=2f_\mathrm{R}T_2^*
\end{equation}
with $f_\mathrm{R}$ the Rabi frequency of the qubit. This quality factor is nothing else than the number of $\pi$ rotations that can be achieved within the dephasing time $T_2^*$. \YMN{Other metrics, involving for example the Rabi dephasing time $T_2^\mathrm{Rabi}$ (relevant for a continuously driven qubit), or the fidelity of single and two-qubit gates may give a more accurate picture of the performances of the device, but require the introduction of extra device-specific parameters (e.g., the complete noise spectrum) unsuitable for a general discussion.}

\YMN{Sweet spots may result from an overall decoupling of the hole and electric fields that comes along with a concomitant reduction of $f_\mathrm{R}$, thus (at best) a constant $Q_2^*$. Such sweet spots may be used to protect the qubit against dephasing while being idle \cite{Bosco21, michal2022tunable}, at the cost of tuning the qubit to a different bias point for manipulation (if possible). Also, an enhancement of $T_2^*$ is \emph{a priori} always beneficial for two qubit gates. Yet the Rabi ``hot spots'' (maximum $f_\mathrm{R}$) of a given gate usually lie near a dephasing sweet spot (and, in particular, along the sweet lines) of that gate owing to ``reciprocal sweetness relations'' between the longitudinal and transverse response of the spin discussed in Ref.~\cite{michal2022tunable}. It is hence possible to enhance both the dephasing time and the Rabi frequency, thus to maximize $Q_2^*$.} Indeed, the Rabi frequency can also be related to the derivative $g^\prime$ of the $g$ matrix with respect to the driving gate voltage \cite{Kato03,Crippa18,Venitucci18}:
\begin{equation}
f_\mathrm{R}=\frac{\mu_BBV_\mathrm{ac}}{2hg^*}\left|(g\mathbf{b})\times( g^\prime\mathbf{b})\right|\,,
\label{eq:fR}
\end{equation}
with $V_{\mathrm{ac}}$ being the amplitude of the drive. Therefore, whenever $f_\mathrm{L}^\prime$ is zero because $g\vec{b}$ is orthogonal to $g^\prime\vec{b}$ (see section \ref{sec:swclass}), these vectors are best oriented for Rabi oscillations. We emphasize that Eq.~\eqref{eq:fR} captures the contributions from both the coupling between the motion/deformation of the dot and the Zeeman Hamiltonian (the so-called $g$-tensor modulation resonance or $g$-TMR \cite{Kato03}), and the effects of the Rashba/Dresselhaus SOC \cite{Venitucci18,Abadillo2023}.

\section{Application to germanium heterostructures}
\label{sec:applications}

\begin{figure}[t]
\begin{center}
\includegraphics[width=0.75\linewidth]{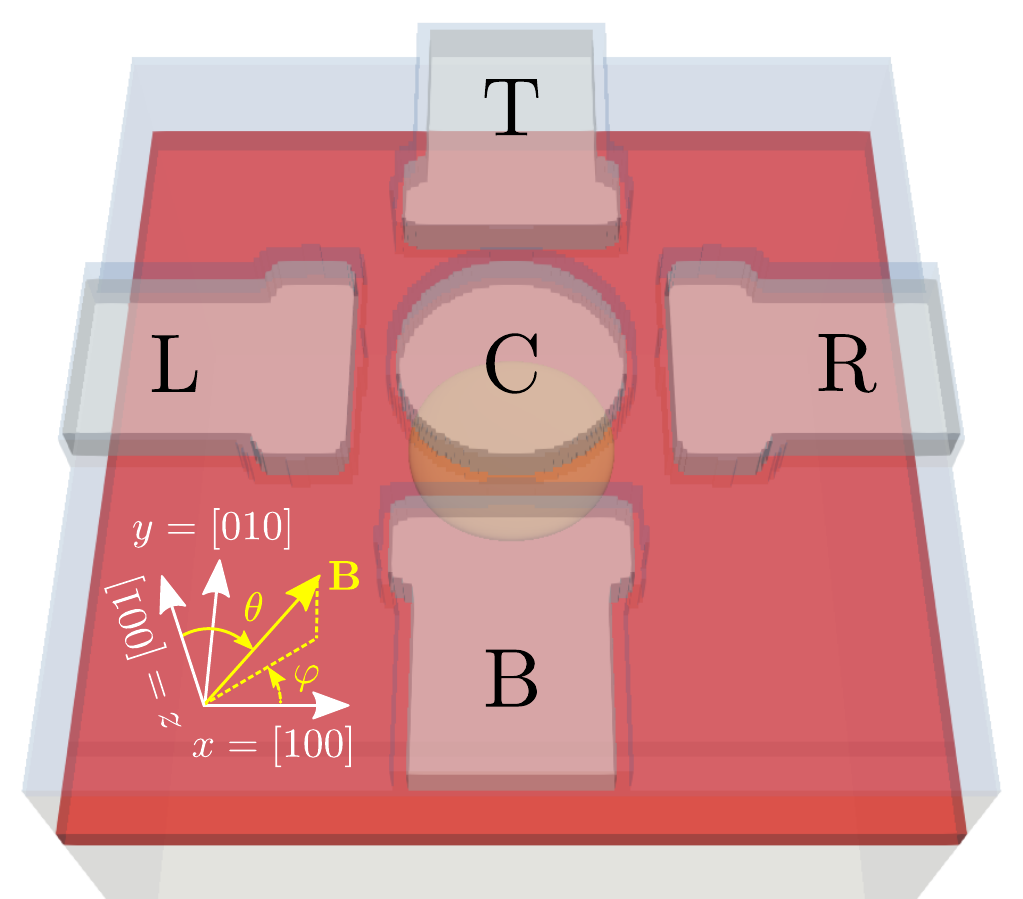}
\caption{Sketch of the device used to simulate a single hole spin qubit in a GeSi heterostructure. The 16\,nm thick Ge quantum well (red) is grown on a thick Ge$_{0.8}$Si$_{0.2}$ buffer and topped with a 50\,nm  thick Ge$_{0.8}$Si$_{0.2}$ barrier (blue). The dot is shaped by the five Al gates C/L/R/T/B (gray). They are 20\,nm
thick and embedded in 5\,nm of Al$_2$O$_3$. The yellow contour is the isodensity surface that encloses 90\% of the ground-state hole charge at bias $V_\mathrm{C}=-40$\,mV (side gates grounded). The orientation of the magnetic field $\vec{B}$ is characterized by the angles $\theta$ and $\varphi$ in the crystallographic axes set $x=[100]$, $y=[010]$ and $z=[001]$.}
\end{center}
\label{fig:device}
\end{figure}

We now apply the above considerations to a hole spin qubit in a Ge/GeSi heterostructure. We consider the prototypical device of Fig.~\ref{fig:device}, identical to Ref.~\cite{Abadillo2023}. It comprises a 16\,nm thick Ge well grown on a Ge$_{0.8}$Si$_{0.2}$ buffer and buried under a 50\,nm thick Ge$_{0.8}$Si$_{0.2}$ barrier. A quantum dot is shaped in the Ge well by the bias voltages applied to the top C/L/R/T/B gates. The diameter of the central C gate is $d=100$\,nm, and the gates are deposited on (and encapsulated in) a 5\,nm thick Al$_2$O$_3$ oxide.

We solve Poisson's equation for the electrostatic potential of the gates with a finite volumes method, then compute the hole wave functions in this potential with a finite-difference discretization of the four bands Luttinger-Kohn Hamiltonian \cite{Luttinger56,KP09}. The latter describes the heavy holes (HH) and light holes (LH) manifold and accounts for the effect of the magnetic field on the orbital and spin degrees of freedom. We calculate the $g$ matrix of the ground-state doublet and its derivatives with respect to the gate voltages along the lines of Ref.~\cite{Venitucci18}, then the Zeeman tensor $G=g^\mathrm{T}g$ and its derivatives $G^\prime=(g^\prime)^\mathrm{T}g+g^\mathrm{T}(g^\prime)$.

We assume a residual in-plane strain $\varepsilon_{xx}=\varepsilon_{yy}=0.26\%$ in the Ge$_{0.8}$Si$_{0.2}$ buffer \cite{Sammak19,Abadillo2023}. The biaxial strains in the Ge well are, therefore, $\varepsilon_{xx}=\varepsilon_{yy}=-0.61\%$ and $\varepsilon_{zz}=+0.45\%$. We may, additionally, account for the inhomogeneous strains transferred to the heterostructure by the thermal contraction of the metal gates upon cool down. The spatial extent of these strains is commensurate with the dot and they can give rise to strong spin-orbit interactions \cite{Abadillo2023}. They are computed with a finite-elements discretization of the continuum elasticity equations. All material parameters (Luttinger parameters, elastic constants...) can be found in Ref. \cite{Abadillo2023}.

In the following, we discuss the impact of the symmetry of the dot on the location of the sweet spots. Therefore, we consider two paradigmatic cases: {\it i}) A highly symmetric circular (CR) quantum dot at bias $V_\mathrm{C}=-40$\,mV with all side gates grounded. The radius of this dot (in homogeneous strains) is $r_\parallel=\sqrt{\langle x^2\rangle+\langle y^2\rangle}=27$\,nm; {\it ii}) A ``squeezed'' dot (SQD) reshaped by side gates biases $V_\mathrm{L}=-V_\mathrm{R}=2$\,mV, $V_\mathrm{T}=95$\,mV and $V_\mathrm{B}=70$\,mV. The extensions of this dot are $\ell_x=\sqrt{\langle x^2\rangle-\langle x\rangle^2}=23$\,nm and $\ell_y=\sqrt{\langle y^2\rangle-\langle y\rangle^2}=14$\,nm, and it is shifted from the center of the C gate by $\langle x\rangle=+8$\,nm and $\langle y\rangle=-6$\,nm. The size of the SQD dot is similar in inhomogeneous strains, but it dot moves further to the right ($\langle x\rangle=+13$\,nm) as the potential along the major $x$ axis is pretty shallow.

We first analyze the $g$-factors of these two dots, then the structure of the sweet spots.

\subsection{$g$-factors}

As discussed in Refs.~\cite{martinez2022hole,Abadillo2023}, the $g$-factors of a HH quantum dot strongly confined along $z=[001]$ are:
\begin{subequations}
\label{eq:gHH}
\begin{align}
g_x&\approx+3q+\frac{6}{m_0\Delta_\mathrm{LH}}\left(\lambda\langle p_x^2\rangle-\lambda^\prime\langle p_y^2\rangle\right) \\
g_y&\approx-3q-\frac{6}{m_0\Delta_\mathrm{LH}}\left(\lambda\langle p_y^2\rangle-\lambda^\prime\langle p_x^2\rangle\right) \\
g_z&\approx6\kappa+\frac{27}{2}q-2\gamma_h+\delta g_z\,.
\end{align}
\end{subequations}
Here $\lambda=\kappa\gamma_2-2\eta_h\gamma_3^2\approx -12.2$ and $\lambda^\prime=\kappa\gamma_2-2\eta_h\gamma_2\gamma_3\approx -5.3$, with $\kappa=3.41$ and $q=0.06$ the isotropic and cubic Zeeman parameters, $\gamma_1=13.38$, $\gamma_2=4.24$ and $\gamma_3=5.69$ the Luttinger parameters of bulk Ge, and $\Delta_\mathrm{LH}\approx 70$\,meV the HH/LH bandgap. $\gamma_h\approx 2.62$ and $\eta_h\approx 0.41$ are factors that depend on vertical confinement and describe the action of the magnetic vector potential on the orbital motion of the holes \cite{Michal21}. $\delta g_z$ is a correction of order $\langle p_x^2\rangle/\Delta_\mathrm{LH}$ and $\langle p_y^2\rangle/\Delta_\mathrm{LH}$ \cite{Katsaros11}. The expectation values of the squared momentum operators $p_x^2$ and $p_y^2$ are computed for the ground-state HH envelope. These expressions account for the effects of the HH/LH mixing by lateral confinement and magnetic field to first order in $1/\Delta_\mathrm{LH}$. We have assumed $\langle p_xp_y\rangle=\langle p_xp_z\rangle=\langle p_yp_z\rangle=0$.

The gyromagnetic factors of the HH ground-state are therefore expected to be highly anisotropic, with $g_x\sim -g_y\sim 3q=0.18$, and $g_z\sim 16$. In the CR dot, where $\langle p_x^2\rangle=\langle p_y^2\rangle$, HH/LH mixing by magnetic confinement slightly decreases $|g_x|$ and $|g_y|$ since $\lambda-\lambda^\prime<0$ \cite{Wang22}. We indeed compute $g_x=-g_y=0.13$ and $g_z=13.7$ in homogeneous strains from the numerical $g$ matrix of the CR dot (hence to all orders in $1/\Delta_\mathrm{LH}$). The gyromagnetic anisotropy is enhanced in germanium by the large $\kappa$ and the small HH/LH mixings. The latter are indeed limited by the wide HH/LH bandgap $\Delta_\mathrm{LH}$ in compressive biaxial strains \cite{Sammak19,martinez2022hole,Abadillo2023}. The $g$-factors are typically much more isotropic in (unstrained) silicon quantum dots \cite{Crippa18,Liles20,Michal21,Piot22}. Squeezing the dot along $y$ shall, according to Eq.~\eqref{eq:gHH}, increase $g_x$ and decrease $g_y$. We actually get principal $g$-factors $g_{x^\prime}=0.26$ and $g_{y^\prime}=0.03$ in the SQD dot. The dot is, therefore, squeezed enough to change the sign of $g_{y^\prime}$ (which has thus a zero at a nearby bias point). The principal magnetic axes $y^\prime\approx x$, $x^\prime\approx y$ and $z^\prime\approx z$ are, moreover, slightly rotated by an angle $\delta\theta=0.16^\circ$ around $\vec{u}\approx\vec{x}$ as the dot does not occupy a symmetric position any more \footnote{This results, as for $g_{zx}^\prime$ and $g_{zy}^\prime$ in Eq.~\eqref{eq:GLprime}, from the non-separability of the in-plane and vertical confinements giving rise to non-zero $g_{zy}\propto\langle p_yp_z\rangle$ and $g_{zx}\propto\langle p_xp_z\rangle$ \cite{martinez2022hole}.}. The principal $g$-factors of the SQD dot are roughly comparable to those reported in Ref.~\cite{Hendrickx2023} (although it is not possible to tell whether $g_{y^\prime}$ has actually changed sign in this experiment). The $g$-factors of the CR and SQD dots are almost the same in inhomogeneous as in homogeneous strains, but the principal axes of the SQD dot rotate around a different axis $\vec{u}\approx(\vec{x}+\vec{y})/\sqrt{2}$. This results from the action of the shear strains, which enhance non-diagonal elements in the $g$ matrix \cite{Abadillo2023}.

\subsection{Sweet lines}

\begin{figure*}[t]
\begin{center}
\includegraphics[width=1.0\linewidth]{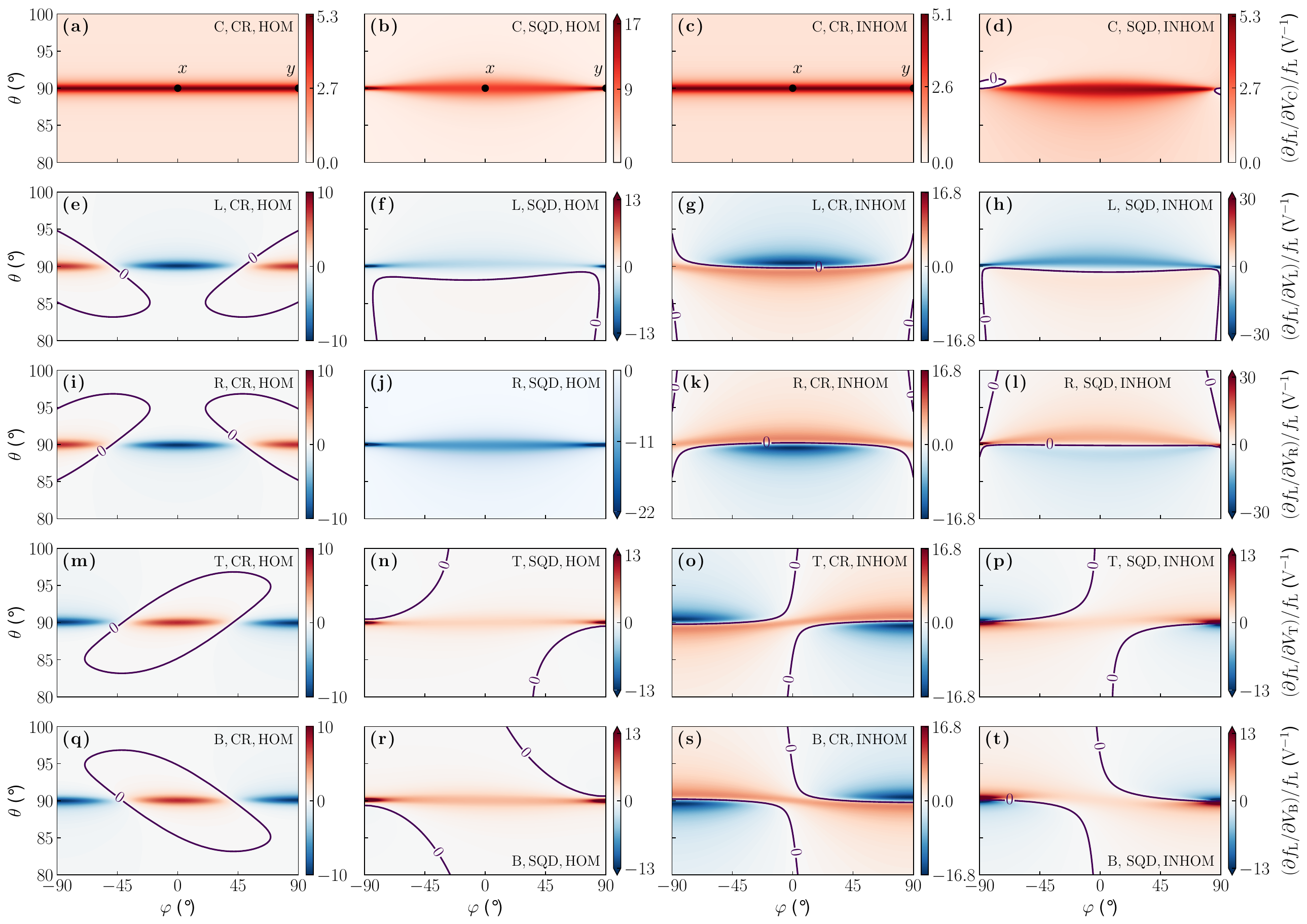}
\caption{The normalized LSES $(\partial f_\mathrm{L}/\partial V_\alpha)/f_\mathrm{L}$, for all gates $\alpha\in\{\mathrm{C},\,\mathrm{L},\,\mathrm{R},\,\mathrm{T},\,\mathrm{B}\}$, in circular (CR) and squeezed (SQD) dots with homogeneous (HOM) and inhomogeneous (INHOM) strains. The sweet spots are highlighted by the purple lines. A map over half the unit sphere $0\le\theta\le 180^\circ$, $-90^\circ\le\varphi\le 90^\circ$ is given in the supplemental material \cite{SI}.}
\label{fig:LSES}
\end{center}
\end{figure*}

The maps of the normalized LSES $(\partial f_\mathrm{L}/\partial V_\alpha)/f_\mathrm{L}$ are plotted as a function of the orientation of the magnetic field in Fig.~\ref{fig:LSES}, for all gates $\alpha\in\{\mathrm{C},\,\mathrm{L},\,\mathrm{R},\,\mathrm{T},\,\mathrm{B}\}$, and for the circular and squeezed dots in homogeneous (HOM) as well as inhomogeneous (INHOM) strains. As both $f_\mathrm{L}$ and $\partial f_\mathrm{L}/\partial V_\alpha$ are proportional to $B$, this normalized LSES is independent on the magnetic field strength, and is the relevant quantity when the device is operated at constant Larmor frequency, as is usually the case. Since Ge spin qubits are typically run with in-plane magnetic fields (where the hole best decouples from hyperfine noise \cite{fischer2008spin,testelin2009hole,Bosco21c} and the Rabi frequency is maximum \cite{martinez2022hole,Abadillo2023,Rodriguez2023}), the normalized LSES is plotted over the range $80^\circ\le\theta\le 100^\circ$, $-90^\circ\le\varphi\le 90^\circ$. The maps over half the unit sphere $0\le\theta\le 180^\circ$, $-90^\circ\le\varphi\le 90^\circ$ can be found in the supplemental material \cite{SI}. The other half of the unit sphere follows from the invariance by the transformation $\vec{B}\to-\vec{B}$.

These maps typically show either no sweet spots or two sweet lines, whose position is, however, highly dependent on the symmetry of the devices and on the SOC mechanisms at work. We analyze these features first for the circular dot, then for the squeezed dot.

\subsubsection{Circular dot}

The central and side gates act differently on the device. Noise on the C gate does not break any symmetry and results in the ``breathing'' of the dot (modulations of the in-plane radius $r_\parallel$ with weaker modulations of the vertical confinement). Although the electric field created by the C gate is essentially vertical \cite{martinez2022hole}, the dot is actually much more sensitive to the smaller in-plane component given its large radius $r_\parallel$. Practically, charge trap fluctuators at the GeSi/Al$_2$O$_3$ interface below the C gate shall induce such a breathing of the dot. The LSES of the side gates describe on the other hand the effects of lateral electric field fluctuations that break the symmetry of the dot (only leaving one mirror plane). The maps of the L and R gates are the same, and differ from those of the B and T gates by a rotation of $90^\circ$ around $z$, as expected from the device layout.

As shown on Fig.~\ref{fig:LSES}, there are no sweet spots with respect to fluctuations on the C gate, and two sweet lines with respect to fluctuations on the side gates. The sign of the eigenvalues of $G^\prime$ in Table \ref{tab:sw} are, therefore, either ``$(+,\,+,\,-)$'' or ``$(-,\,-,\,+)$'' for the side gates. 

As a matter of fact, the high symmetry of the dot constrains the shape of the matrices $g_\alpha^\prime=\partial g/\partial V_\alpha$ and $G_\alpha^\prime=\partial G/\partial V_\alpha$ \cite{Venitucci18}. The Bloch functions of the heavy holes can be mapped onto the $J_z=\pm\tfrac{3}{2}$ components of a $J=3/2$ angular momentum; choosing $\ket{\Uparrow}\equiv\ket{+\tfrac{3}{2}}$ and $\ket{\Downarrow}\equiv\ket{-\tfrac{3}{2}}$ as the states with main $J_z=\pm\tfrac{3}{2}$ character respectively, the $g$ matrix reads in the principal magnetic axes $x$, $y$, $z$:
\begin{equation}
g=
\begin{pmatrix}
g_x & 0 & 0 \\
0 & g_y & 0 \\
0 & 0 & g_z  
\end{pmatrix}\,.
\end{equation}
with $g_x=-g_y$ and $g_z$ given (to lowest order in $1/\Delta_\mathrm{LH}$) by Eqs. (\ref{eq:gHH}). For the C gate on the one hand, $g^\prime$ then takes the form \cite{martinez2022hole}
\begin{equation}
g_\mathrm{C}^\prime=
\begin{pmatrix}
g_x^\prime & 0 & 0 \\
0 & g_y^\prime & 0 \\
0 & 0 & g_z^\prime  
\end{pmatrix}
\Rightarrow
G_\mathrm{C}^\prime=2
\begin{pmatrix}
g_x g_x^\prime & 0 & 0 \\
0 & g_y g_y^\prime & 0 \\
0 & 0 & g_z g_z^\prime 
\end{pmatrix}\,,
\end{equation}
while for the left (or right) gate on the other hand,
\begin{align}
g_\mathrm{L}^\prime=
\begin{pmatrix}
g_x^\prime & 0 & g_{xz}^\prime \\
0 & g_y^\prime & 0 \\
g_{zx}^\prime & 0 & g_z^\prime  
\end{pmatrix}
\Rightarrow
G_\mathrm{L}^\prime=2
\begin{pmatrix}
g_x g_x^\prime & 0 & \tfrac{1}{2}G_{xz}^\prime \\
0 & g_y g_y^\prime & 0 \\
\tfrac{1}{2}G_{zx}^\prime & 0 & g_z g_z^\prime 
\end{pmatrix}\,,
\label{eq:GLprime}
\end{align}
with $G_{xz}^\prime=G_{zx}^\prime=g_x g_{xz}^\prime+g_z g_{zx}^\prime$. The matrices for the B and T gates are similar, with instead non-zero $g_{yz}^\prime$, $g_{zy}^\prime$, and $G_{yz}^\prime=G_{zy}^\prime=g_y g_{yz}^\prime+g_z g_{zy}^\prime$. In the following, we label $G_x^\prime=2g_x g_x^\prime$, $G_y^\prime=2g_y g_y^\prime$ and $G_z^\prime=2g_z g_z^\prime$ the diagonal elements of the $G^\prime$ matrices.

The matrix $G_\mathrm{C}^\prime$ is purely diagonal, which means that the C gate modulates the principal $g$-factors but not the principal magnetic axes. All the diagonal elements of $G_\mathrm{C}^\prime$ take the same positive sign when the dot breathes in the electric field of the C gate ($G_x^\prime=G_y^\prime=0.18$\,V$^{-1}$, $G_z^\prime=174.3$\,V$^{-1}$). There is, therefore, no sweet spot with respect to fluctuations of the C gate voltage. In principle, $G_z^\prime$ may change sign at negative enough $V_\mathrm{C}=V_\mathrm{C,sw}$, when the increase of the overlap between the HH and LH envelopes that enhances cubic Rahsba SOC is compensated by the opening of the HH/LH bandgap \cite{Wang21}. This would bring two sweet points along $z$ at $V_\mathrm{C}=V_\mathrm{C,sw}$, then two sweet lines swirling around $z$ at $V_\mathrm{C}<V_\mathrm{C,sw}$. However, the vertical electric field needed to reach this sweet spot is presumably of the order of $20$\,mV/nm \cite{Wang21,Sarkar23}, far beyond the maximum $\approx 3$\,mV/nm that the Ge well can sustain before the hole gets pulled out by the electric field \cite{martinez2022hole}. We can therefore conclude that there are no sweet spots with respect to vertical electric field fluctuations and breathing in the operational gate voltages range. The normalized LSES of the C gate is maximal for in-plane magnetic fields because $g_x^\prime/g_x$ and $g_y^\prime/g_y$ are particularly sensitive to breathing, as highlighted by Eqs.~\eqref{eq:gHH}. Even though $G_z^\prime=g_z g_z^\prime$ is the largest diagonal element of $G^\prime$, $g_z$ actually shows the smallest relative variations $g_z^\prime/g_z=G_z^\prime/(2g_z^2)$. Moreover, $g_z$ rapidly dominates the Zeeman splitting once the magnetic field goes out of plane, as $g_z/|g_{x,y}|\approx 100$. This gives rise to a sharp in-plane peak in the normalized LSES with full width at half maximum
\begin{equation}
\delta b_z=2\left(1+\frac{g_z^2}{g_x^2}-2\frac{G_z^\prime}{G_x^\prime}\right)^{-1/2}\approx 2\left|\frac{g_x}{g_z}\right|\approx 0.021\,,
\label{eq:deltabz}
\end{equation}
or equivalently $\delta\theta\approx 1.2^\circ$.

The electric field from a side gate (or a side defect) deforms the dot and breaks the disk-shape symmetry. As a result, $G_x^\prime$ and $G_y^\prime$ are of opposite sign according to Eqs.~\eqref{eq:gHH}, which gives rise to two sweet lines crossing the equatorial plane at four points. For the L and R gate, $G_x^\prime=-0.33$\,V$^{-1}$, $G_y^\prime=0.25$\,V$^{-1}$ and $G_z^\prime=-35.8$\,V$^{-1}$, so that the sweet lines swirl around $y$ if $G_{xz}^\prime=0$. Correspondingly, the sweet lines swirl around $x$ for the B and T gates if $G_{yz}^\prime=0$. The sweet lines are however tilted by the non-zero $G_{xz}^\prime$ and $G_{yz}^\prime$ that track down the rotations of the principal magnetic axes [see Eqs.~\eqref{eq:gammaxi}]. The components of $\vec{b}=(b_x,\,b_y,\,b_z)$ must for example fulfill on the sweet line of the L or R gate:
\begin{equation}
G_x^\prime b_x^2+G_y^\prime b_y^2+G_z^\prime b_z^2+2G_{xz}^\prime b_xb_z=0\,.
\label{eq:swL}
\end{equation}
\YMN{The cubic Rashba SOC \cite{Rashba88,Winkler03,Marcellina17,Wang21,Terrazos21} gives rise to non-zero but opposite $g_xg_{xz}^\prime\approx -g_zg_{zx}^\prime$ and $g_yg_{yz}^\prime\approx -g_zg_{zy}^\prime$ \cite{martinez2022hole,Abadillo2023}, thus to negligible $G_{xz}^\prime$ and $G_{yz}^\prime$}. In homogeneous strains, $g_{zx}^\prime$ and $g_{zy}^\prime$ essentially result from the coupling between the in-plane and vertical motions of the hole in the non-separable potential of the gates (namely, this potential is not the simple sum of an homogeneous electric field along $z$ and of a harmonic confinement along $x$ and $y$) \cite{martinez2022hole}. The dot thus ``rocks'' out-of-plane while moving in-plane, which gives rise to a rotation $\{x,\,y,\,z\}\to\{x^\prime,\,y^\prime,\,z^\prime\}$ of the principal magnetic axes around $x$ ($g_{zy}^\prime\ne 0$) or $y$ ($g_{zx}^\prime\ne 0$). The effect of this rotation on the Zeeman splitting is strongly amplified by the large $g_z$ ($G_{xz}^\prime\approx g_z g_{zx}^\prime$ and $G_{yz}^\prime\approx g_z g_{zy}^\prime$), because small excursions of the magnetic field around the principal $(x^\prime y^\prime)$ plane can result in large variations of $g_z^2 b_{z^\prime}^2$ (with respect to $g_x^2b_{x^\prime}^2+g_y^2b_{y^\prime}^2\sim g_{x,y}^2$). The magnitude of $G_{xz}^\prime$ and $G_{yz}^\prime$ remains, however, ten times smaller than $|G_z^\prime|$. In inhomogeneous strains, $g_{zx}^\prime$ and $g_{zy}^\prime$ pick an additional contribution from the motion of the hole in the gradients of shear strains $\varepsilon_{xz}$ and $\varepsilon_{yz}$ that modulate the HH/LH mixings \cite{Abadillo2023}. \YMNN{$G_{xz}^\prime=G_{yz}^\prime=45.7$\,V$^{-1}$} then outweigh $G_x^\prime$ and $G_y^\prime$ by two orders of magnitude and are comparable to $G_z^\prime$. According to Eq.~\eqref{eq:swL}, the sweet lines of the L and R gates shall be $\vec{b}\perp\vec{x}$ and $\vec{b}\perp\vec{z}$ in the limit $|G_{xz}^\prime|\gg |G_x^\prime|,\,|G_y^\prime|,\,|G_z^\prime|$, while those of the B and T gates shall be $\vec{b}\perp\vec{y}$ and $\vec{b}\perp\vec{z}$ (at large enough $|G_{yz}^\prime|$). In practice, the sweet lines always run close to the equatorial plane in inhomogeneous strains (despite the comparatively large $G_z^\prime$) because $G_{xz}^\prime$ and $G_{yz}^\prime$ have no effect when $b_z=0$ and because $|G_{xz}^\prime|$ (or $|G_{yz}^\prime|$) are much greater than $|G_x^\prime|$ and $|G_y^\prime|$.

As for the C gate, the sensitivity to lateral electric field fluctuations is stronger near in-plane magnetic fields. The sharpness of the in-plane features is again a consequence of the large $g_z/|g_{x,y}|$ ratio \cite{martinez2022hole,Abadillo2023}, which gives rise to fast variations of both $f_\mathrm{L}$ and $f_\mathrm{L}^\prime$ around $b_z=0$. In particular, the LSES rapidly changes sign around the equatorial plane in inhomogeneous strains as a result of the action of $G_{xz}^\prime$ and $G_{yz}^\prime$. Interestingly, $\sqrt{|G_x^\prime/G_y^\prime|}\approx 1$ for all side gates, so that their sweet lines cross the equatorial plane near (but not exactly at) $\vec{b}=(\pm\vec{x}\pm\vec{y})/\sqrt{2}$, in both homogeneous and inhomogeneous strains. 

\subsubsection{Squeezed dot}

The normalized LSES of the SQD dot are quantitatively different from those of the CR dot. First of all, the four side gates now play non equivalent roles, so that their LSES maps are not related any more by symmetry operations. Moreover, squeezing the dot along $y$ (thus increasing $\langle p_y^2\rangle$) increases both $g_x$ and $g_y$ according to Eqs.~\eqref{eq:gHH}. In the present case, the dot is squeezed enough to change the sign of $g_y$ (now positive), hence of $G_y^\prime=2g_yg_y^\prime$ (unless $g_y^\prime$ also changes sign). As a consequence, the R gate map lacks sweet spots in homogeneous strains (where the effects of the off-diagonal element $G_{xz}^\prime$ do not yet prevail). On the other hand, the C gate map now shows two sweet lines around $\approx y$, but only in inhomogeneous strains (because $g_y^\prime$ also changes sign in homogeneous strains \footnote{The squeezed dot does not breath homothetically in the $x$ and $y$ directions and the sign of $g_y^\prime$ becomes highly dependent on the shift $\langle y\rangle$ of the dot with respect to the central axis of the C gate.}). The axes of these sweet lines are shifted from $y$ by the shear strains, which give rise to non-zero $G_{xz}^\prime$, $G_{yz}^\prime$ and $G_{xy}^\prime$ even for the C gate as the squeezed dot has moved with respect to the symmetry axis of the strains (the $z$ axis). Another consequence of the change of sign of $G_y^\prime$ is that there are most often no sweet spots in the equatorial plane any more: the sweet lines do not cross that plane and run slightly above or below. 

The magnitude of the normalized LSES looks much larger in the SQD than in the CR dot, especially around $y$ (except for the C gate in inhomogeneous strains). Indeed, the denominator of the normalized LSES is $\propto (g^*)^2$ and is thus very small along that direction in the SQD dot (namely, a given $g_y^\prime$ results in much larger relative variations of the Larmor frequency in the SQD than in the CR dot). The raw LSES (at constant field) stand on more comparable scales (see supplemental material \cite{SI}). Those of the L and R gates are actually larger in the SQD than in the CR dot, while those of the T and B gates are smaller, because the SQD dot is more responsive to electric fields along its major than along its minor axis. Nonetheless, the LSES of the CR and SQD dots are qualitatively similar in inhomogeneous strains where the physics is dominated by the large $G_{xz}^\prime$ and $G_{yz}^\prime$ brought by the shear strains $\varepsilon_{xz}$ and $\varepsilon_{yz}$. 

\subsubsection{Discussion}

The above analysis demonstrates how the LSES of the gates highlight general properties of the dephasing rates and electrical sweet spots of Ge/GeSi spin qubits. First of all, the LSES of all gates show strong variations near the equatorial plane owing to the large ratio between $g_z$ and $g_{x,y}$, which reverses the balance between the $\propto B_{x,y}$ and $\propto B_z$ components of the Zeeman splitting over a degree around that plane. Moreover, any change of the shape of the hole wave function results in large relative variations of the small $g_x$ and $g_y$, which are strongly dependent on the orbital motion of the holes (the $\propto 1/\Delta_\mathrm{LH}$ correction in Eqs.~\eqref{eq:gHH}). The normalized LSES of the C gate peaks within the equatorial plane and usually shows no exploitable sweet spots. This underlines that the dot can hardly be protected against breathing in a noisy environment, as $|g_x|$ and $|g_y|$ then show similar variations. On the other hand, the LSES of the side gates, which describe the response to lateral electric fields, typically show sweet lines crossing the equatorial plane as $|g_x|$ and $|g_y|$ vary in opposite ways (unless the dot is squeezed enough to change the sign of one in-plane $g$-factor). In inhomogeneous cool down strains, the fluctuations of the Zeeman splitting are ruled by the motion of the dot in the shear strains, which give rise to rotations of the principal magnetic axes. The effects of these rotations are again amplified by the large $|g_z/g_{x,y}|$ ratio. In that regime, the LSES of the side gates are qualitatively less dependent on the symmetry of the dot. They show sweet lines running very close to the equatorial plane $\theta=90^\circ$; however, these features are very sharp and enclosed between nearby ``hot'' lines \YMNN{where the absolute LSES gets maximum. As an example, the splitting between the two hot spots on either side of the sweet line in the $(xz)$ plane ($\varphi=0$) of Figs.~\ref{fig:LSES}g,k is
\begin{equation}
\delta b_z\approx 2\left|\frac{g_x}{g_z}\right|\sqrt{1+\left(\frac{g_x^\prime}{g_{zx}^\prime}\right)^2}\approx 0.020
\end{equation}
or $\delta\theta\approx 1.17^\circ$. This splitting is, like Eq.~(\ref{eq:deltabz}), limited by the small $|g_x/g_z|$, which squeezes sweet and hot features within a very narrow angular range.}

These clusters of sweet and hot spots practically call for a very accurate alignment of the magnetic field. Moreover, there are no common sweet spots between the different side gates, which further complicates the engineering of dephasing times. The variations of the LSES shall be much smoother in systems with more homogeneous principal $g$-factors \YMNN{such as silicon \cite{Piot22} or light-hole \cite{DelVecchio2023} quantum dots}. This shall provide in principle more opportunities for design optimizations aimed at bringing the sweet spots as close as possible to maximize dephasing times, as discussed in Ref. \cite{Piot22}. 

The above analysis also shows that the LSES of the gates can convey a lot of information about the spin-orbit coupling mechanisms prevailing in the device, in lieu of or in complement with the measurement of the Rabi frequencies of the same gates \cite{Crippa19}. The measurement of the LSES is, moreover, easier in principle than the measurement of the Rabi frequencies. The main characteristics of the LSES computed in this work are compatible with the experimental findings of Ref.~\cite{Hendrickx2023}. Yet the LSES of the side gates of Ref.~\cite{Hendrickx2023} reveal additional rotations of the principal magnetic axes in the $(xy)$ plane (evidenced by a large $G_{xy}^\prime)$. These rotations result most likely from the motion of the dot in the non-separable in-plane potential or shear strains $\varepsilon_{xy}$ \cite{Michal21,Abadillo2023}, whose effects are enhanced by the lower symmetry of the gate layout.

\begin{figure}[t]
\begin{center}
\includegraphics[width=1.0\linewidth]{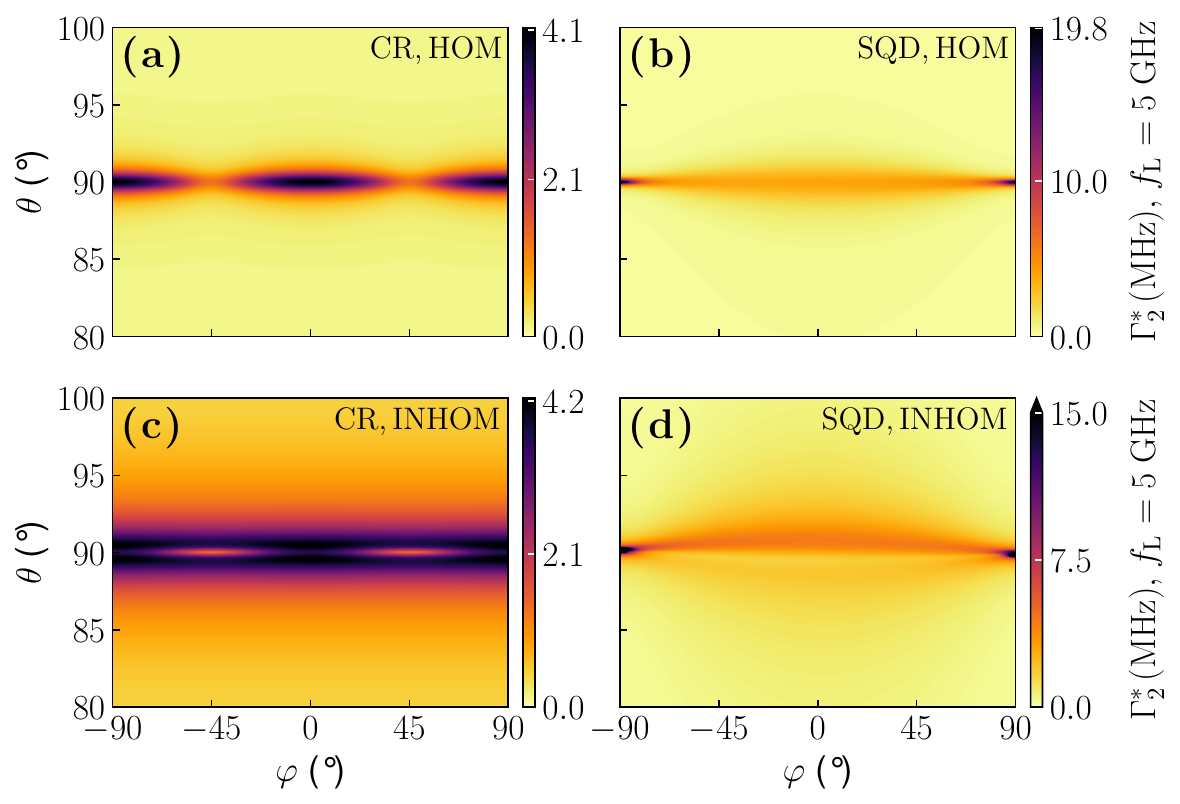}
\caption{Total dephasing rate $\Gamma_2^*$ (MHz) at constant Larmor frequency $f_\mathrm{L}=5$\,GHz in circular (CR) and squeezed (SQD) dots with homogeneous (HOM) and inhomogeneous (INHOM) strains. $\Gamma_2^*$ is computed with Eq.~\eqref{eq:gammatot} assuming the same $\delta V^\mathrm{rms}=10$\,$\mu$V on all gates.}
\label{fig:gammatot}
\end{center}
\end{figure}

In appendix \ref{app:traps}, we show that the LSES of the gates reproduce the main features of the LSES of individual charge traps at the GeSi/Al$_2$O$_3$ interface. We further discuss how far the LSES of the gates provides a comprehensive description of the response to electrical and charge noise. In the following, we address the total dephasing rate and quality factors in a simple approximation.

\subsection{Dephasing times and quality factors}

Our purpose is to draw general (rather than device specific) conclusions about dephasing times and quality factors in Ge/GeSi spin qubits. We hence assume as a simple but illustrative example that dephasing is dominated by direct electrical noise on the gates and by the exchange of carriers between the gates and traps in the oxide below \cite{Shalak23}. We also consider isotopically purified materials and discard hyperfine interactions. In these assumptions, the LSES of the gates shall provide a reasonable description of the response of the qubit (see Appendix \ref{app:traps}). We thus estimate the total dephasing rate $\Gamma_2^*$ with Eq.~\eqref{eq:gammatot}, assuming $1/f$-like noise with the same $\delta V^\mathrm{rms}=10$\,$\mu$V on all gates. $\Gamma_2^*$ is hence directly proportional to $\delta V^\mathrm{rms}$ and $f_\mathrm{L}$.

\begin{figure*}[ht!]
\begin{center}
\includegraphics[width=1.0\linewidth]{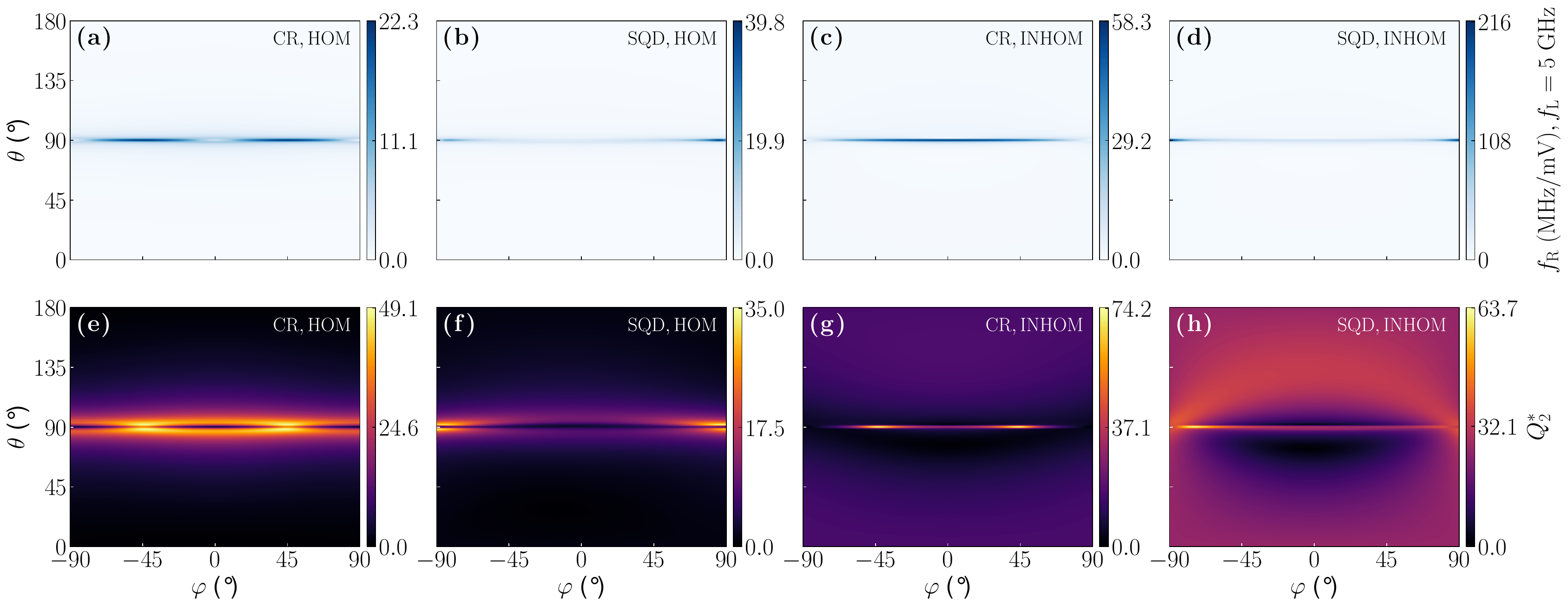}
\caption{(a)-(d) Rabi frequencies $f_\mathrm{R}$ and (e)-(h) quality factor $Q_2^*=2f_\mathrm{R}T_2^*$ when driving with the L gate, for circular (CR) and squeezed (SQD) dots in homogeneous (HOM) and inhomogeneous (INHOM) strains. The Rabi frequencies are calculated at constant Larmor frequency $f_\mathrm{L}=5$\,GHz, and the amplitude of the AC signal on the L gate is $V_\mathrm{ac}=1$\,mV.}
\label{fig:Q}
\end{center}
\end{figure*}

The maps of the resulting $\Gamma_2^*$ are plotted \YMN{around the equatorial plane} in Fig.~\ref{fig:gammatot}, for the CR and SQD devices in both homogeneous and inhomogeneous strains. These maps are calculated at constant Larmor frequency $f_\mathrm{L}=5$\,GHz (maps on the half unit sphere are also provided in the supplemental material \cite{SI}). We recover on these plots the main features highlighted on the LSES of the single gates. The dephasing rate of the circular dot in homogeneous strains peaks in the equatorial plane; yet it shows local optima in that plane near $\varphi=\pm 45^\circ$ that are the fingerprints of the sweet spots of the side gates. These features may not, however, survive if the in-plane electric field noise is more ``isotropic'' (not specifically oriented along the gate axes, see Appendix \ref{app:traps}). In inhomogeneous strains, the dephasing peak gets broader, but is split by a thin dip in the equatorial plane as a result of the structure of the LSES of the side gates, with sweet lines running close to (and crossing) that plane flanked by two peaks. The dephasing rate of the squeezed dot also shows a broad peak around the equatorial plane, and is maximum near $\vec{B}\parallel\vec{y}$ (essentially because the magnetic field must be raised there to keep the Larmor frequency constant). There is no particular structure standing out from $\Gamma_2^*$ in homogenenous strains as the sweet lines of the different gates are too far apart. However, a faint dip does appear on the bottom side of the main peak in inhomogeneous strains as a result of the interplay between the different sweet lines running close to the equatorial plane. Despite very different LSES, the total dephasing rates remain comparable in homogeneous and inhomogeneous strains.

Finally, we plot on Fig.~\ref{fig:Q} the Rabi frequencies $f_\mathrm{R}$ and quality factors $Q_2^*=2f_\mathrm{R}T_2^*$ achieved in all four cases when driving the L gate with an AC signal $V_\mathrm{L}(t)=V_\mathrm{ac}\sin(2\pi f_\mathrm{L}t)$. The Rabi frequencies are computed for $V_\mathrm{ac}=1$\,mV and at constant Larmor frequency $f_\mathrm{L}=5$\,GHz. The quality factors are actually independent on $f_\mathrm{L}$ and are proportional to $V_\mathrm{ac}/\delta V^\mathrm{rms}$ since $f_\mathrm{L}\propto V_\mathrm{ac}B$ but $T_2^*\propto 1/(\delta V^\mathrm{rms}B)$ for $1/f$-noise. The Rabi frequencies peak in-plane owing to the spin-orbit coupling mechanisms at work and to the large $g_z/|g_{x,y}|$ ratio that leads to a strong increase of $f_\mathrm{R}\propto B$ around the equatorial plane \cite{martinez2022hole,Abadillo2023}. The Rabi oscillations for in-plane magnetic fields essentially result from the same mechanisms as the LSES, which couple the motion of the dot to the Zeeman Hamiltonian of the hole and promote $g$-TMR \cite{Kato03}. For the circular dot in homogeneous strain, they are primarily driven by the deformations of the dot in the inhomogeneous AC electric field of the L gate, \YMN{which give rise to non-zero $g_x^\prime$ and $g_y^\prime$ and to the two peaks at $\varphi\approx\pm 45^\circ$ \cite{martinez2022hole}.} Owing to the reciprocal sweetness relations discussed in section \ref{sec:recriprocity}, these Rabi ``hot spots'' lie close to the sweet lines of the left gate (see Fig.~\ref{fig:LSES}e). This remains true in inhomogeneous strains, even though the Rabi oscillations are now primarily driven by the motion of the dot in the shear strains $\varepsilon_{xz}$ \YMN{(resulting in non-zero $g_{zx}^\prime$)} \cite{Abadillo2023}. The small oscillations of the principal magnetic axes are indeed converted in an efficient $g$-tensor modulation by the large $g_z$. In the squeezed dot, the maximum Rabi frequencies are shifted toward the $y$ axis. This results from the small $g_y$, which again leads to a strong enhancement of $f_\mathrm{R}\propto B$ at constant Larmor frequency. While squeezing the dot gives rise to the emergence of linear Rashba spin-orbit interactions \cite{Bosco21b}, they remain too small to help spin manipulation. Driving with the C gate is also possible, but usually slower (see supplemental material \cite{SI}) \cite{Martinez2022,Abadillo2023}; the Rabi frequency maps are similar (up to a rotation) for the R/T/B gates in the circular dot, but the T and B gates are less efficient in the squeezed dot that is little polarizable along its minor axis $y$.

The quality factor $Q_2^*$ also peaks near the equatorial plane, even though both $f_\mathrm{R}$ and $\Gamma_2^*$ decrease out of this plane. This can be understood from the dependence of $f_\mathrm{R}$ on $b_z$ as given by Eq.~\eqref{eq:fR} with Eq.~\eqref{eq:GLprime} for $g_\mathrm{L}^\prime$. For magnetic fields in the $(xz)$ plane for example,
\begin{equation}
f_\mathrm{R}\propto\frac{1}{g^*}\left|(g_zg_x^\prime-g_xg_z^\prime)b_xb_z+g_zg_{xz}^\prime b_z^2-g_xg_{zx}^\prime b_x^2\right|\,.
\end{equation}
\YMN{Therefore, $f_\mathrm{R}\propto |g_{xz}^\prime b_z|$ at large $|b_z|\gg |b_x|$, while $\Gamma_2^*\propto |{\bar g}_z^\prime b_z|$  with $({\bar g}_z^\prime)^2$ the average $(g_z^\prime)^2$ of all gates, so that $Q_2^*\propto |g_{xz}^\prime/{\bar g}_z^\prime|$. Whereas ${\bar g}_z^\prime$ picks a significant contribution from the deformation of the wave function, $g_{xz}^\prime$ exclusively results from cubic Rashba SOC \cite{martinez2022hole} in homogeneous strains. The latter is, however, little efficient at low vertical electric fields \cite{Marcellina17,Wang21,Terrazos21}, and thus $Q_2^*$ is small at large $b_z$.} This means that the Larmor vector $\boldsymbol{\Omega}=g\vec{B}$ gets locked on the $z$ axis and shows stronger longitudinal (LSES) than transverse (Rabi) susceptibility. The situation is more favorable in inhomogeneous strains, which give rise to a linear Rashba contribution to $g_{xz}^\prime$ that enhances the out-of-plane Rabi frequencies and quality factors \cite{Abadillo2023}. In homogeneous strains, the best quality factors are achieved along two lines running slightly above and below the equatorial plane as a result of the competition between faster Rabi oscillations but also larger dephasing rates when approaching that plane. In inhomogeneous strains, these lines merge because the large $g_{zx}^\prime$ and $g_{zy}^\prime$ brought by shear strains make purely transverse contributions when $b_z=0$ (that is, give rise to fast Rabi oscillations but do not contribute to the LSES). This is also why the maximal in-plane quality factors are typically larger in inhomogeneous than in homogeneous strains \cite{Abadillo2023}. This in-plane feature is however very thin as the same $g_{zx}^\prime$ and $g_{zy}^\prime$ start to make large longitudinal contributions (described by $G_{xz}^\prime$ and $G_{yz}^\prime$) once the magnetic field goes out-of-plane. It is also slightly shifted away from the equatorial plane in the squeezed dot owing to the small rotation of the principal axes pre-existing in the undriven dot. The Rabi hot spots/dephasing sweet spots near $\varphi\approx\pm 45^\circ$ are clearly visible in homogeneous strains.

These data suggest that the optimal point operation of Ge/GeSi hole spin qubits shall be sought in or near the equatorial plane, but is expected to be very sharp. \YMN{This may give rise to a large variability that will complicate the management of arrays of spin qubits, especially when the principal axes get misaligned by electric fields (deforming the dots) and strains \YMNN{\cite{Wang2024}}, and/or when the dominant source of noise differs from device to device. Although hole spin qubits are highly tunable in principle, there are practically limited margins for adjustment as the same gates are used to confine the holes and drive one- as well as two-qubit operations. It can, therefore, be difficult to find a single magnetic field orientation that will be suitable for all qubits (e.g., where the performance metrics are at least half of the optimum). In Ge/SiGe heterostructures, out-of-plane quality factors may still be significant (and more uniform) in inhomogeneous strains, at the price of a much slower manipulation. More generally, the management of sweet spots in large arrays of qubits shall be easier in materials with more homogeneous $g$-factors, such as silicon, where the sweet features are expected broader \cite{Piot22}. However, holes at the Si/SiO$_2$ interface suffer from stronger scattering by the charged defects in the amorphous SiO$_2$ \cite{Martinez2022}. The $g$-factor anisotropy in Ge/SiGe may also be alleviated by materials and device engineering, using, e.g., uniaxial stressors to increase $|\varepsilon_{xx}-\varepsilon_{yy}|$ (thus $|g_x+g_y|$) \cite{Abadillo2023} or, on the opposite, working with weakly strained Ge wells to enhance the HH/LH mixing.}

\section{Conclusions}

We have investigated the dephasing times and sweet spots of a spin-orbit qubit electrically coupled to a fluctuator. For that purpose, we have characterized the Larmor frequency $f_\mathrm{L}=\mu_B\sqrt{\vec{B}^\mathrm{T}G\vec{B}}/h$ of this qubit by a Zeeman tensor $G$, and the coupling to the fluctuator by its derivative $G^\prime$ with respect to the fluctuating field. We have shown that the geometrical nature of the sweet spots on the unit sphere describing the magnetic field orientation depends on the sign of the eigenvalues of $G^\prime$. In most cases, the sweet spots of a single fluctuator draw a couple of lines on this sphere (rather than simple points), which affords more opportunities for optimal operation of the qubit. We have also discussed how the measurement of the derivatives $G^\prime$ with respect to gate voltages gives valuable insights into spin-orbit coupling, and provides a meaningful characterization of the sensitivity of the device to electrical and charge noise. We have then applied this framework to a model hole spin qubit in a Ge/GeSi heterostructure. We have analyzed the arrangement of the sweet lines of the different gates, and their relations with the physics of the device (spin-orbit coupling mechanisms at work, symmetry of the dot, gyromagnetic factors anisotropy, ...). The sweet structures in Ge/GeSi heterostructures appear very thin owing to the strong anisotropy of the $g$-factors, which calls for a careful alignment of the magnetic field and may be a significant source of device-to-device variability in arrays of spin qubits. The engineering of sweet spots and dephasing times shall be easier in materials with a smaller anisotropy thus broader sweet features. \YMNN{This work actually provides the tools to explore the relevance of design optimizations aiming at extending coherence. }\YMN{Finally, we would like to point out that the $g$-matrix formalism used in this work to compute Rabi frequencies and dephasing rates is not only a powerful tool for single qubits, but finds natural extensions to two-qubit systems, as illustrated in Ref.~\cite{Sen2023}.}

\section*{Acknowledgements}

This work is supported by the French National Re-
search Agency under the programme ``France 2030'' (PEPR
PRESQUILE - ANR-22-PETQ-0002), by the European
Union’s Horizon 2020 research and innovation program
(Grant Agreement No. 951852 QLSI) and European Research
Council (ERC) Project No. 810504 (QuCube).

\appendix

\section{LSES of charge traps at the GeSi/Al$_2$O$_3$ interfaces}
\label{app:traps}

In this appendix, we discuss how far the LSES of the gates provide a faithful description of the response of the quantum dot to noise.

\begin{figure*}[t]
\begin{center}
\includegraphics[width=1.0\linewidth]{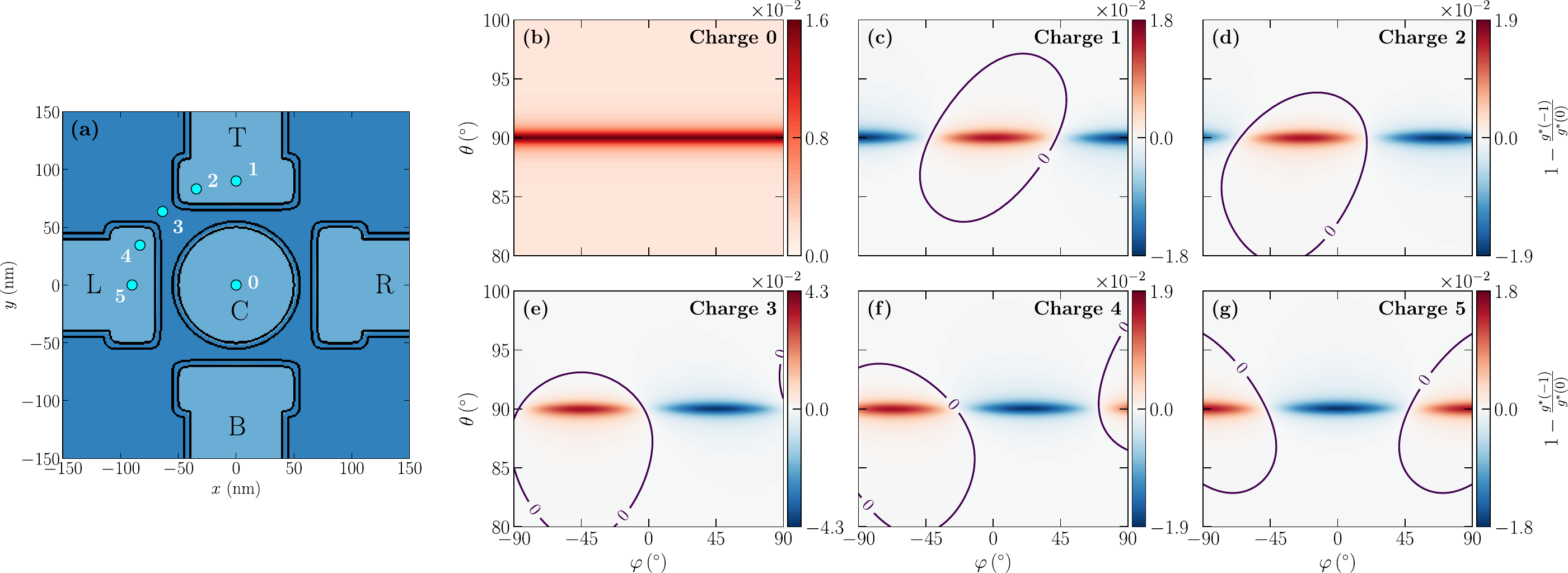}
\caption{(b)-(g) LSES of single traps at the different locations displayed in (a), for the CR dot in homogeneous strains. The LSES of single traps is defined by Eq.~\eqref{eq:LSESq}.}
\label{fig:traps}
\end{center}
\end{figure*}

For that purpose, we compute the normalized LSES of single charge traps at the GeSi/Al$_2$O$_3$ interface, defined as:
\begin{align}
\frac{1}{f_\mathrm{L}(0)}\frac{\partial f_\mathrm{L}}{\partial q}&=\frac{1}{f_\mathrm{L}(0)}\left[f_\mathrm{L}(0)-f_\mathrm{L}(-1)\right] \nonumber \\
&=1-\frac{g^*(-1)}{g^*(0)}\,,
\label{eq:LSESq}
\end{align}
where $f_\mathrm{L}(q)$ and $g^*(q)$ are the Larmor frequency and gyromagnetic factor when the trap holds a charge $q$. We consider negative charge traps, but the results are similar for positive ones. If we model charge capture/release as a random telegraph noise with switching rate $\nu_\mathrm{t}$ \cite{Shalak23}, the coherence decays with an exponential envelope $\exp(-\Gamma_2^* t)$ \cite{Abel08,Bergli09,Paladino14}, where:
\begin{equation}
\Gamma_2^*=\nu_\mathrm{t}\left[1-\mathrm{Re}\sqrt{1-\left(\frac{\pi}{\nu_\mathrm{t}}\frac{\partial f_\mathrm{L}}{\partial q}\right)^2}\right]\,.
\end{equation}
In the limit $\nu_\mathrm{t}\gg\pi\partial f_\mathrm{L}/\partial q$ -- namely, when the charge fluctuator is fast with respect to the Larmor frequency shift, this expression simplifies to:
\begin{equation}
\Gamma_2^*\approx\frac{\pi^2}{2\nu_\mathrm{t}}\left(\frac{\partial f_\mathrm{L}}{\partial q}\right)^2\,.
\label{eq:G2imp}
\end{equation}
The dephasing rate then decreases as $1/\nu_\mathrm{t}$ as the Larmor frequency behaves as a Gaussian random variable whose fluctuations progressively average out on time scale $(\partial f_\mathrm{L}/\partial q)^{-1}$. On the other hand, in the opposite limit $\nu_\mathrm{t}<\pi\partial f_\mathrm{L}/\partial q$ (slow fluctuator), $\Gamma_2^*$ gets upper bounded by $\nu_\mathrm{t}$ because dephasing can not be faster than the average time between two switches of the trap. We emphasize that the Gaussian (fast fluctuator) regime always holds around the sweet spots of a given trap. $\Gamma_2^*$ scales as $B^2$ in this regime, as expected from Bloch-Redfield theory (which reproduces Eq.~\eqref{eq:G2imp}) \cite{Paladino14}. The transition to $1/f$-like noise is believed to result from the interplay between many traps with switching rates $\nu_\mathrm{t}$ distributed as $1/\nu_\mathrm{t}$ \cite{Bergli09}.

The LSES of traps located at six different positions are plotted on Fig.~\ref{fig:traps} for the circular dot in homogeneous strains. The LSES of trap 0, located along the axis of the C gate, and the LSES of traps 1 and 5, located along the axes of the L and T gates, resemble those of the corresponding gate in Fig.~\ref{fig:LSES}. They feature approximately the same sweet lines with two sweet points near $\varphi=\pm 45^\circ$ in the equatorial plane. The LSES of traps 2, 3 and 4, which occupy less symmetric positions around the dot, are qualitatively similar to that of trap 1, but appear rotated by the same angle as their radius vector makes with $y$. The sweet points in the equatorial plane are, in particular, shifted away from $\varphi\approx\pm 45^\circ$. Although this could be expected from the high symmetry of the quantum dot, similar conclusions hold for the squeezed dot (in both homogeneous and inhomogeneous strains): the LSES of the impurities resemble those of the gates but are rotated according to the impurity position around the dot.

This underlines that the LSES of the gates may, practically, not be able to probe all relevant perturbations. In particular, none of the gates can give rise to (and probe) a large $G_{xy}^\prime$ in the present layout. If the most limiting charge fluctuator(s) are located below the gates (where they can easily exchange carriers with the latter) \cite{Shalak23}, then the LSES of the gates shall capture the most important features of the dephasing rates. If, on the contrary, the most limiting fluctuator(s) are located between the gates (where they are presumably slower, but unscreened), the dephasing rates may appear rotated by 45$^\circ$ with respect to the LSES of the gates. The orientation of the minimum in-plane dephasing rate is typically orthogonal to the dominant electric field noise. However, for a broad distribution of traps without strongly dominant fluctuator(s), the angular dependence of Figs.~\ref{fig:traps}b-f will likely be averaged out, so that the dephasing rates shall look like Fig.~\ref{fig:traps}b (or equivalently like the LSES of the C gate).

In the supplementary material of Ref.~\cite{Abadillo2023}, we have discussed the LSES of the circular dot with respect to joint gate voltage modulations $\delta V_\mathrm{R}=-\delta V_\mathrm{L}$. Such modulations indeed break exactly the same symmetries as a homogeneous in-plane electric field oriented along $x$, a standard test perturbation in simple models for quantum dot spin qubits. This LSES is nothing else than the sum of panels (e) and (i), or (g) and (k) of Fig.~\ref{fig:LSES}. It shows two sweet lines $b_z=0$ (in-plane magnetic fields) and $b_x=0$, because $g_x^\prime=g_y^\prime=g_z^\prime=0$ for perturbations with this symmetry \cite{Venitucci18,martinez2022hole}. However, such joint modulations seem little representative of the actual noise in real devices with uncorrelated charge fluctuators all around the qubits. We have, therefore, discarded joint modulations in the present study.

\bibliography{biblio}

\clearpage

\setcounter{section}{0}
\setcounter{equation}{0}
\setcounter{figure}{0}
\setcounter{table}{0}

\renewcommand\thefigure{S\arabic{figure}} 
\renewcommand\theequation{S\arabic{equation}}

\onecolumngrid

\begin{center}
\textbf{\large Supplemental material for ``Geometry of the dephasing sweet spots of spin-orbit qubits''}
\end{center}

\begin{figure*}[b]
\begin{center}
\includegraphics[width=1.0\linewidth]{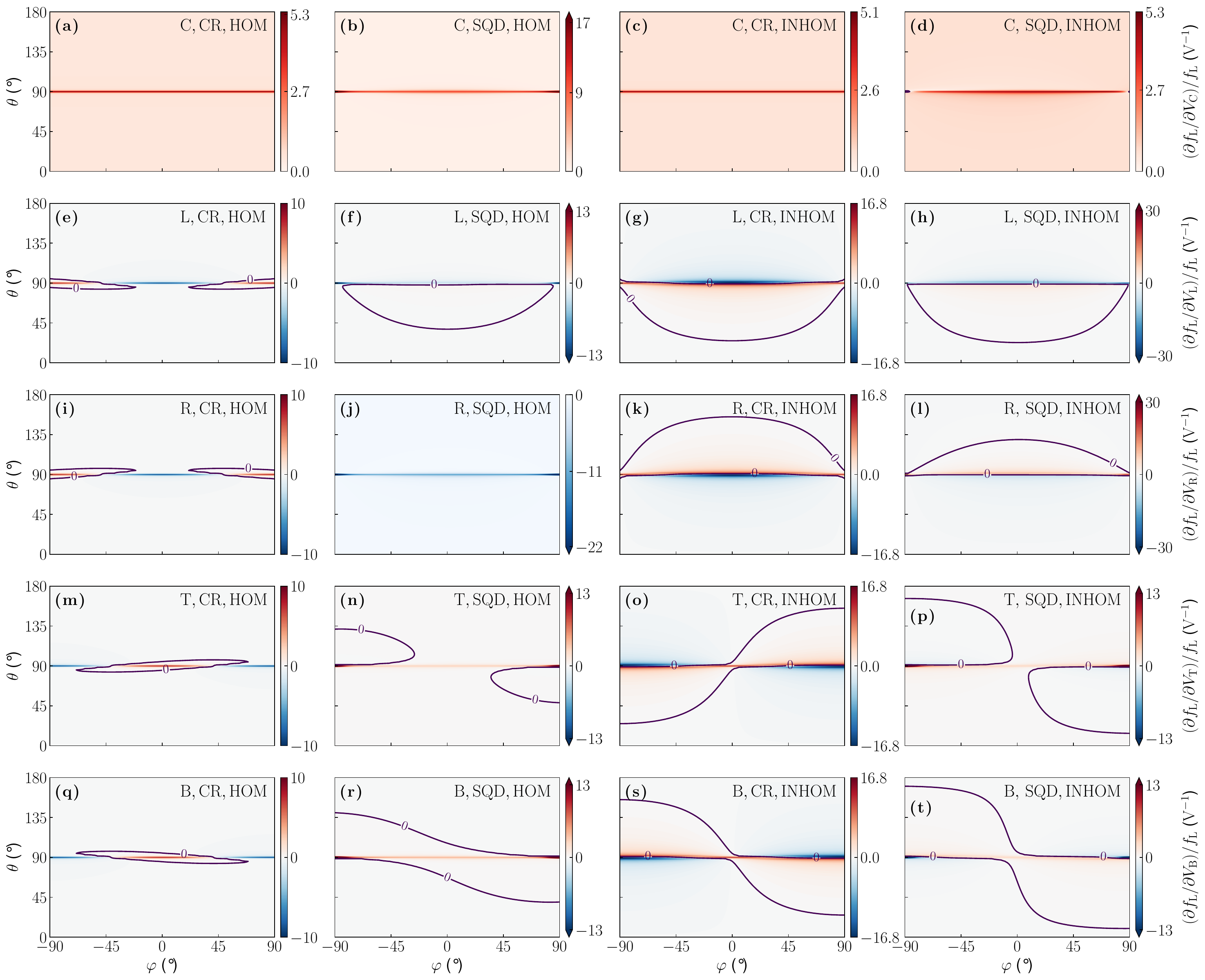}
\caption{The normalized LSES $(\partial f_\mathrm{L}/\partial V_\alpha)/f_\mathrm{L}$, for all gates $\alpha\in\{\mathrm{C},\,\mathrm{L},\,\mathrm{R},\,\mathrm{T},\,\mathrm{B}\}$, in circular (CR) and squeezed (SQD) dots with homogeneous (HOM) and inhomogeneous (INHOM) strains. The sweet spots are highlighted by the purple lines. This figure is the same as Fig. 2 of the main text, but plotted over half the unit sphere. The other half of the unit sphere follows from the invariance by the transformation $\vec{b}\to-\vec{b}$.}
\label{fig:LSES_FullRange}
\end{center}
\end{figure*}

\begin{figure*}[t]
\begin{center}
\includegraphics[width=1.0\linewidth]{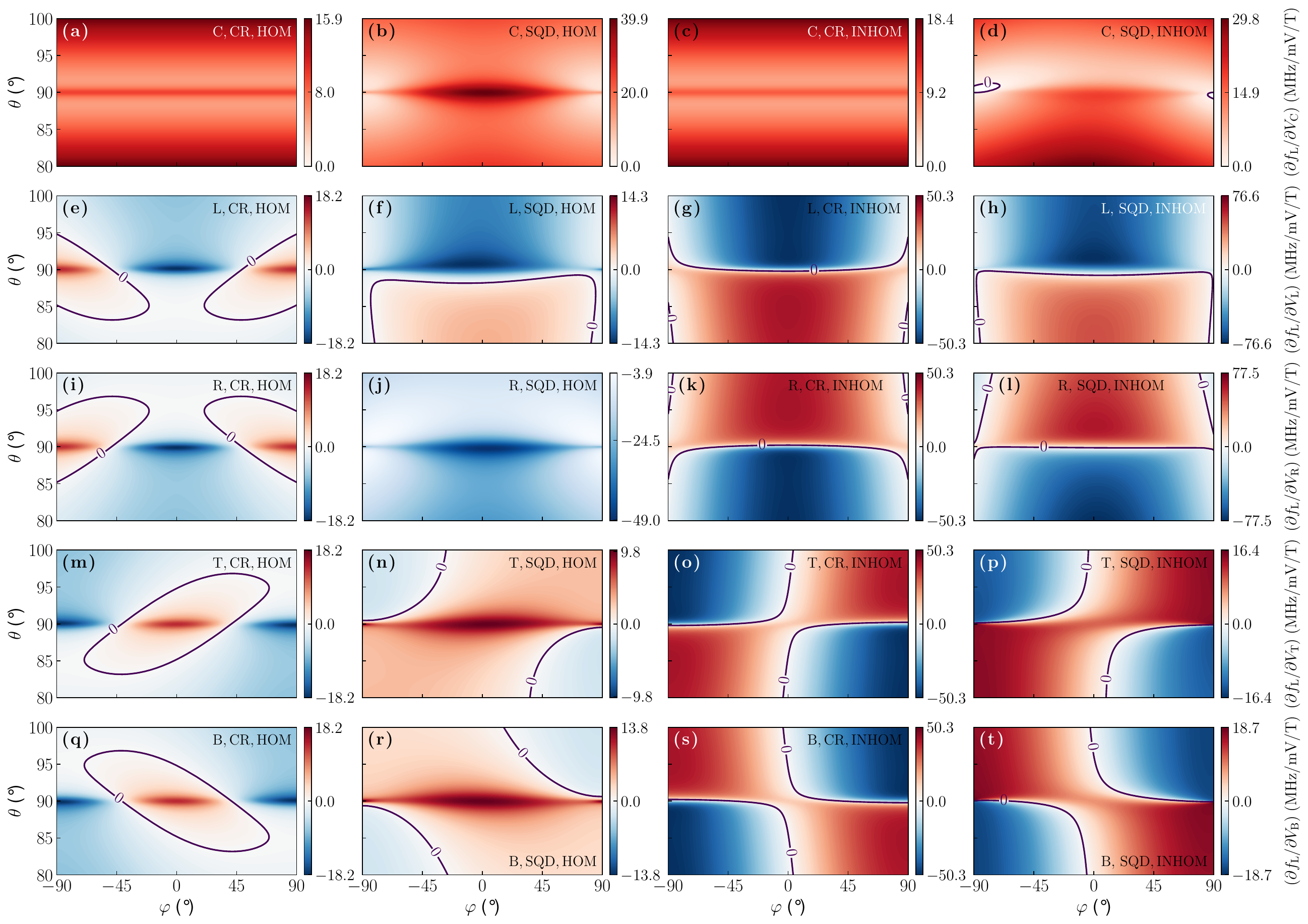}
\caption{The raw LSES $(\partial f_\mathrm{L}/\partial V_\alpha)$, for all gates $\alpha\in\{\mathrm{C},\,\mathrm{L},\,\mathrm{R},\,\mathrm{T},\,\mathrm{B}\}$, in circular (CR) and squeezed (SQD) dots with homogeneous (HOM) and inhomogeneous (INHOM) strains. The sweet spots are highlighted by the purple lines. The raw LSES characterizes dephasing at constant magnetic field amplitude $B$, and is thus given in MHz/mV/T. These maps are zooms over $80\le\theta\le 100^\circ$, $-90^\circ\le\varphi\le 90^\circ$.}
\label{fig:LSES_noNorm}
\end{center}
\end{figure*}

\begin{figure*}[t]
\begin{center}
\includegraphics[width=1.0\linewidth]{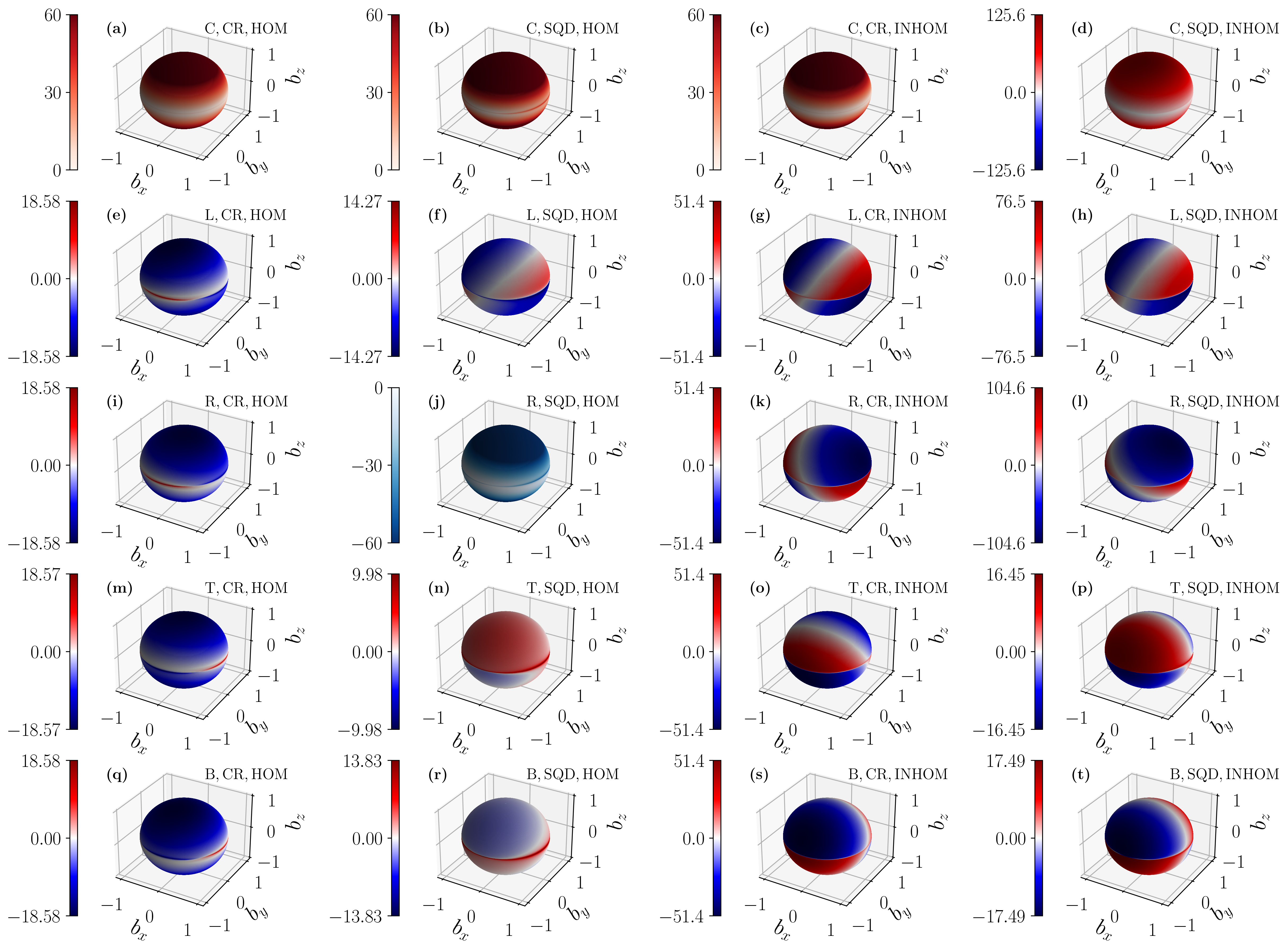}
\caption{The raw LSES $(\partial f_\mathrm{L}/\partial V_\alpha)$ on the unit sphere, for all gates $\alpha\in\{\mathrm{C},\,\mathrm{L},\,\mathrm{R},\,\mathrm{T},\,\mathrm{B}\}$, in circular (CR) and squeezed (SQD) dots with homogeneous (HOM) and inhomogeneous (INHOM) strains. This figure is the same as Fig.~\ref{fig:LSES_noNorm}, plotted over the unit sphere rather than as a 2D map (which deforms the sweet lines).}
\label{fig:LSES_FullRange_Sphere}
\end{center}
\end{figure*}

\begin{figure*}[t]
\begin{center}
\includegraphics[width=0.66\linewidth]{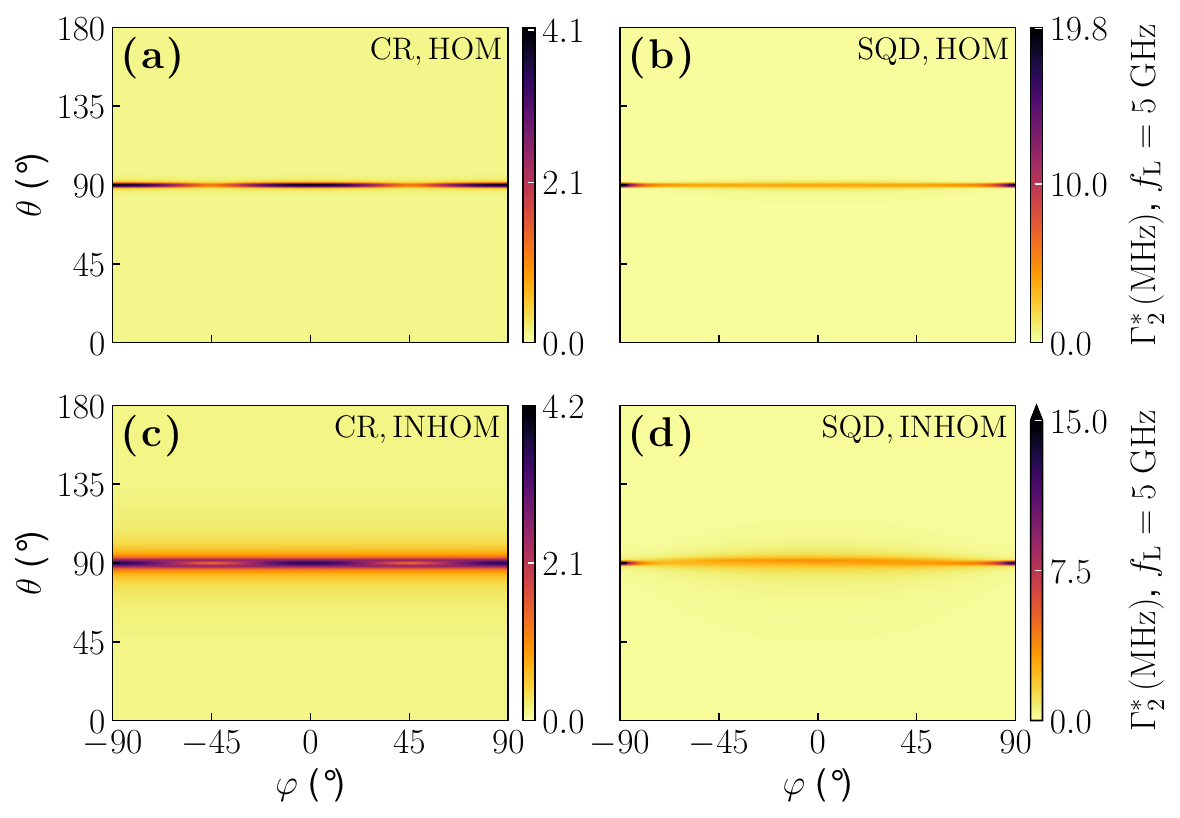}
\caption{Total dephasing rate $\Gamma_2^*$ (MHz) at constant Larmor frequency $f_\mathrm{L}=5$\,GHz in circular (CR) and squeezed (SQD) dots with homogeneous (HOM) and inhomogeneous (INHOM) strains. $\Gamma_2^*$ is computed with Eq. (21) of the main text assuming the same $\delta V^\mathrm{rms}=10$\,$\mu$V on all gates. This figure is the same as Fig. 3 of the main text, but plotted over half the unit sphere.}
\label{fig:Gamma_theta0180}
\end{center}
\end{figure*}

\begin{figure*}[t]
\begin{center}
\includegraphics[width=1.0\linewidth]{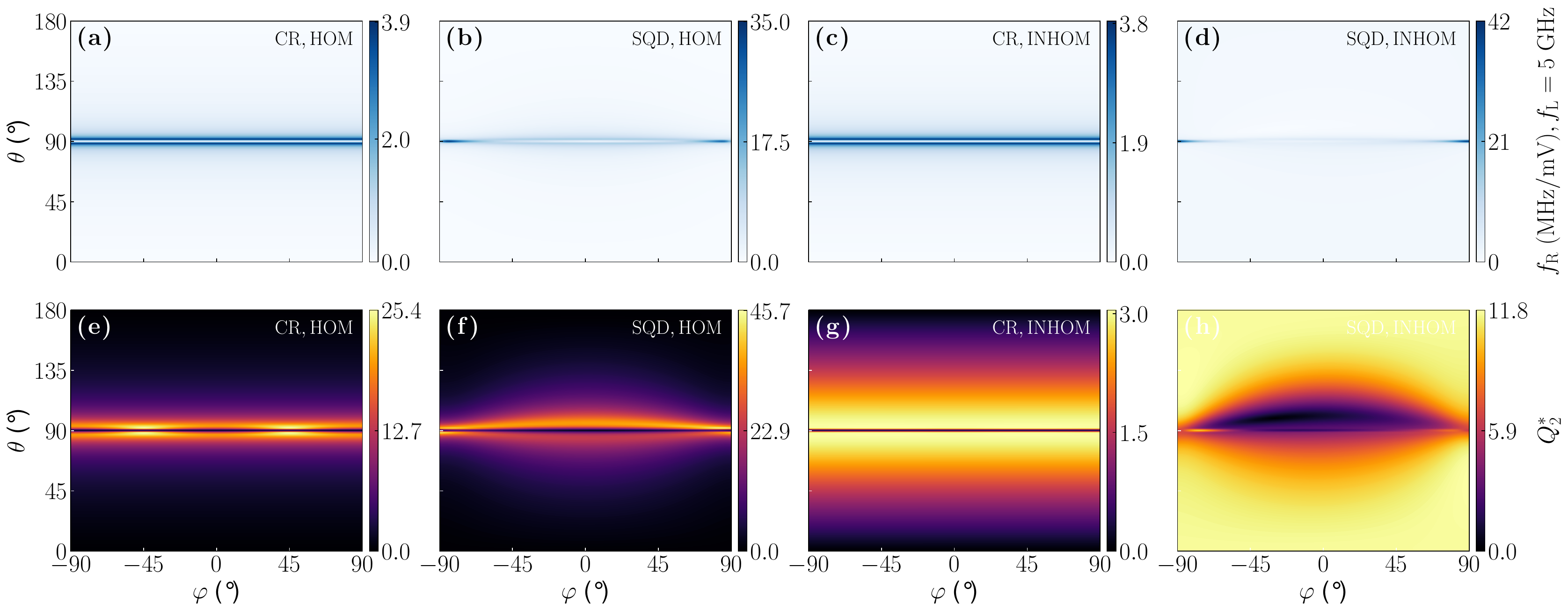}
\caption{(a)-(d) Rabi frequencies $f_\mathrm{R}$ and (e)-(h) quality factor $Q_2^*=2f_\mathrm{R}T_2^*$ when driving with the C gate, for circular (CR) and squeezed (SQD) dots in homogeneous (HOM) and inhomogeneous (INHOM) strains. The Rabi frequencies are calculated at constant Larmor frequency $f_\mathrm{L}=5$\,GHz, and the amplitude of the AC signal on the C gate is $V_\mathrm{ac}=1$\,mV.}
\label{fig:QualityFactors_Cdriving}
\end{center}
\end{figure*}

In this supplemental material, we provide:
\begin{itemize}
    \item Maps of the normalized LSES $(\partial f_\mathrm{L}/\partial V_\alpha)/f_\mathrm{L}$ of the gates over half the unit sphere (Fig.~\ref{fig:LSES_FullRange}), as a supplement to Fig. 2 of the main text,
    \item Maps of the raw LSES $\partial f_\mathrm{L}/\partial V_\alpha$ of the gates at constant magnetic field amplitude $B$ (Figs.~\ref{fig:LSES_noNorm} and \ref{fig:LSES_FullRange_Sphere}),
    \item Maps of the dephasing rate $\Gamma_2^*$ over half the unit sphere (Fig.~\ref{fig:Gamma_theta0180}), as a supplement to Fig. 3 of the main text.
    \item Maps of the Rabi frequency $f_\mathrm{R}$ and quality factor $Q_2^*$ (Fig.~\ref{fig:QualityFactors_Cdriving}) when driving with the C gate, as a supplement to Fig. 4 of the main text.
\end{itemize}

\end{document}